\documentclass[superscriptaddress,aps,prl,showkeys,showpacs,twocolumn,10pt]{revtex4}
\usepackage[OT1,T1]{fontenc}
\usepackage{graphicx}
\usepackage{rotating}
\begin{document}

\title{Neutron-Proton Scattering in the Context of the $d^*$(2380) Resonance}
\date{\today}

\newcommand*{\IKPUU}{Division of Nuclear Physics, Department of Physics and 
 Astronomy, Uppsala University, Box 516, 75120 Uppsala, Sweden}
\newcommand*{\ASWarsN}{Department of Nuclear Physics, National Centre for 
 Nuclear Research, ul.\ Hoza~69, 00-681, Warsaw, Poland}
\newcommand*{\IPJ}{Institute of Physics, Jagiellonian University, ul.\ 
 Reymonta~4, 30-059 Krak\'{o}w, Poland}
\newcommand*{\PITue}{Physikalisches Institut, Eberhard--Karls--Universit\"at 
 T\"ubingen, Auf der Morgenstelle~14, 72076 T\"ubingen, Germany}
\newcommand*{\Kepler}{Kepler Center for Astro and Particle Physics, University
 of T\"ubingen, Auf der Morgenstelle~14, 72076 T\"ubingen, Germany}
\newcommand*{\MS}{Institut f\"ur Kernphysik, Westf\"alische 
 Wilhelms--Universit\"at M\"unster, Wilhelm--Klemm--Str.~9, 48149 M\"unster, 
 Germany}
\newcommand*{\ASWarsH}{High Energy Physics Department, National Centre for 
 Nuclear Research, ul.\ Hoza~69, 00-681, Warsaw, Poland}
\newcommand*{\IITB}{Department of Physics, Indian Institute of Technology 
 Bombay, Powai, Mumbai--400076, Maharashtra, India}
\newcommand*{\PGI}{Peter Gr\"unberg Institut, Forschungszentrum J\"ulich,
  52425 J\"ulich, Germany}
\newcommand*{\ILP}{Institut f\"ur Laser- und Plasmaphysik, Heinrich-Heine
  Universit\"at D\"usseldorf, 40225 D\"usseldorf, Germany}
\newcommand*{\IKPJ}{Institut f\"ur Kernphysik, Forschungszentrum J\"ulich, 
 52425 J\"ulich, Germany}
\newcommand*{\JCHP}{J\"ulich Center for Hadron Physics, Forschungszentrum 
 J\"ulich, 52425 J\"ulich, Germany}
\newcommand*{\Bochum}{Institut f\"ur Experimentalphysik I, Ruhr--Universit\"at 
 Bochum, Universit\"atsstr.~150, 44780 Bochum, Germany}
\newcommand*{\ZELJ}{Zentralinstitut f\"ur Engineering, Elektronik und 
 Analytik, Forschungszentrum J\"ulich, 52425 J\"ulich, Germany}
\newcommand*{\Erl}{Physikalisches Institut, 
 Friedrich--Alexander--Universit\"at Erlangen--N\"urnberg, 
 Erwin--Rommel-Str.~1, 91058 Erlangen, Germany}
\newcommand*{\ITEP}{Institute for Theoretical and Experimental Physics, State 
 Scientific Center of the Russian Federation, Bolshaya Cheremushkinskaya~25, 
 117218 Moscow, Russia}
\newcommand*{\Giess}{II.\ Physikalisches Institut, 
 Justus--Liebig--Universit\"at Gie{\ss}en, Heinrich--Buff--Ring~16, 35392 
 Giessen, Germany}
\newcommand*{\IITI}{Department of Physics, Indian Institute of Technology 
 Indore, Khandwa Road, Indore--452017, Madhya Pradesh, India}
\newcommand*{\Aachen}{III.~Physikalisches Institut~B, Physikzentrum, 
 RWTH Aachen, 52056 Aachen, Germany}
\newcommand*{\HepGat}{High Energy Physics Division, Petersburg Nuclear Physics 
 Institute, Orlova Rosha~2, Gatchina, Leningrad district 188300, Russia}
\newcommand*{\HeJINR}{Veksler and Baldin Laboratory of High Energiy Physics, 
 Joint Institute for Nuclear Physics, Joliot--Curie~6, 141980 Dubna, Russia}
\newcommand*{\Katow}{August Che{\l}kowski Institute of Physics, University of 
 Silesia, Uniwersytecka~4, 40-007, Katowice, Poland}
\newcommand*{\Tomsk}{Department of Physics, Tomsk State University,
634050 Tomsk, Russia}
\newcommand*{\IFJ}{The Henryk Niewodnicza{\'n}ski Institute of Nuclear 
 Physics, Polish Academy of Sciences, 152~Radzikowskiego St, 31-342 
 Krak\'{o}w, Poland}
\newcommand*{\NuJINR}{Dzhelepov Laboratory of Nuclear Problems, Joint 
 Institute for Nuclear Physics, Joliot--Curie~6, 141980 Dubna, Russia}
\newcommand*{\KEK}{High Energy Accelerator Research Organisation KEK, Tsukuba, 
 Ibaraki 305--0801, Japan} 
\newcommand*{\ASLodz}{Department of Cosmic Ray Physics, National Centre for 
 Nuclear Research, ul.\ Uniwersytecka~5, 90--950 {\L}\'{o}d\'{z}, Poland}

\author{P.~Adlarson}    \affiliation{\IKPUU}
\author{W.~Augustyniak} \affiliation{\ASWarsN}
\author{W.~Bardan}      \affiliation{\IPJ}
\author{M.~Bashkanov}   \affiliation{\PITue}\affiliation{\Kepler}
\author{F.S.~Bergmann}  \affiliation{\MS}
\author{M.~Ber{\l}owski}\affiliation{\ASWarsH}
\author{H.~Bhatt}       \affiliation{\IITB}
\author{M.~B\"uscher}\affiliation{\PGI}\affiliation{\ILP}
\author{H.~Cal\'{e}n}   \affiliation{\IKPUU}
\author{I.~Ciepa{\l}}   \affiliation{\IPJ}
% ...

\newcommand*{\Delhi}{Department of Physics and Astrophysics, University of 
 Delhi, Delhi--110007, India}
\newcommand*{\Mainz}{Institut f\"ur Kernphysik, Johannes 
 Gutenberg--Universit\"at Mainz, Johann--Joachim--Becher Weg~45, 55128 Mainz, 
 Germany}
\newcommand*{\Bern}{Albert Einstein Center for Fundamental Physics, University 
 of Bern, Sidlerstrasse~5, 3012 Bern, Switzerland}
\newcommand*{\Sheff}{Department of Physics and Astronomy, University of 
 Sheffield, Hounsfield Road, Sheffield, S3 7RH, United Kingdom}

\collaboration{WASA-at-COSY Collaboration}\noaffiliation
\newcommand*{\GW}{Data Analysis Center at the Institute for Nuclear Studies,
  Department of Physics, The George Washington University, Washington,
  D.C. 20052, U.S.A.}
\author{R. L. Workman}     \affiliation{\GW}
\author{W. J. Briscoe}     \affiliation{\GW}
\author{I. I. Strakovsky}  \affiliation{\GW}
\collaboration{SAID Data Analysis Center}

\begin{abstract}
New data on quasifree polarized neutron-proton scattering, in the region of the
recently observed $d^*$ resonance structure, have been obtained by exclusive and
kinematically complete high-statistics measurements with WASA at COSY.
This paper details the determination of the beam
polarization, checks of the quasifree character of the scattering process,
on all obtained $A_y$ angular distributions and on the new partial-wave
analysis, 
which includes the new data producing a resonance pole in the $^3D_3$-$^3G_3$
coupled partial waves at ($2380\pm10 - i40\pm5$) MeV -- in accordance with the
$d^*$ dibaryon resonance hypothesis. The effect of the new 
partial-wave solution on the description of total and differential cross
section data as well as specific combinations of spin-correlation and
spin-transfer observables available from COSY-ANKE measurements at $T_d$ =
2.27 GeV is discussed. 

\end{abstract}


\maketitle

\section{Introduction}

Recent measurements of the basic double-pionic fusion to the deuteron, which
comprises the reaction channels $pn \to d\pi^0\pi^0$, $pn \to d\pi^+\pi^-$ and
$pp \to d\pi^+\pi^0$, reveal a narrow resonance-like structure in the total
cross section at a mass $M \approx$ 2380 MeV with a width of $\Gamma \approx$
70 MeV \cite{mb,MB,isofus}. From the isospin decomposition of the cross
sections in the three fusion channels the isoscalar nature of this
structure has been determined \cite{isofus}, whereas the determination of its
spin-parity $J^P = 3^+$ has been obtained from the angular distributions in
the $d\pi^0\pi^0$ channel, which has a particularly low background from   
conventional reaction processes \cite{MB}. Further support for this resonance
structure has been found in the $pn \to pp\pi^0\pi^-$ reaction \cite{TS},
where it was denoted by $d^*$ -- following its notation associated with the
so-called "inevitable" dibaryon \cite{goldman,Mulders} having identical quantum
numbers. 

If the observed resonance-like structure constitutes an $s$-channel 
resonance in the neutron-proton system, then it has to be sensed also in the 
observables of elastic $np$ scattering. In Ref. \cite{PBC} this
resonance effect in $np$ scattering has been estimated and it was shown 
that a noticeable effect should appear in the analyzing power $A_y$. This
observable is  most sensitive to small changes in the partial waves, since it
is composed only of interference terms between partial waves.

For the analyzing power, there are data only below and above the resonance
region. These data sets, at $T_n$~=~1.095 GeV ($\sqrt s$ = 2.36 GeV)
\cite{ball,les} and $T_n$~=~1.27 GeV ($\sqrt s$ = 2.43 GeV) \cite{mak,dieb},
exhibit very similar angular distributions. This gap in the existing
measurements of $A_y$ has motivated the present study, the main results of
which have been communicated recently already in a Letter \cite{prl2014}.

\section{Experiment}

The measurements of polarized $\vec{n}p$ elastic scattering over the energy
region of interest have been carried out in the quasifree mode. The
experiment was performed with the WASA detector \cite{CB,wasa} at the
COSY storage ring by use of a polarized deuteron beam of energy of
$T_d$~=~2.27~GeV, which impinged 
on the WASA hydrogen pellet target. Utilizing the quasifree scattering
process the full energy region of the conjectured resonance was
covered. Since we observe here the quasi-free scattering process 
$\vec d p \to np + p_{spectator}$ in inverse kinematics, we were able to
detect the fast spectator proton in the forward detector of WASA.

Since elastic $np$ scattering has a large cross section, it was sufficient to
have a trigger, which solely requested one hit in the first layer of the
forward range hodoscope. Such a hit could originate from either a charged
particle or a neutron. For the case of quasifree $np$ scattering we thus have
three event classes, with each of them having the spectator proton hit the 
forward detector:
\begin{itemize}
\item  (i) Both scattered proton and scattered neutron are detected in the central
  detector. This event type covers the region $31^\circ < \Theta_n^{cm} <
  129^\circ$ of neutron angles. 
\item (ii) The scattered proton is detected in the forward detector, whereas the
  scattered neutron is not measured. This type concerns the region $132^\circ <
  \Theta_n^{cm} < 178^\circ$.
\item (iii) The scattered proton is detected in the central detector with the
  neutron being unmeasured, since its scattering angle is outside the angular
  range of the central detector. This event type covers the angular range
  $30^\circ 
  < \Theta_n^{cm} < 41^\circ$.
\end {itemize}

Thus nearly the full range of neutron scattering angles could be 
covered by use of all three event classes.

Since by use of inverse kinematics the spectator proton resides in the
deuteron beam particle, the emitted spectator proton is very fast, which
facilitates its detection in the forward detector. Thus by reconstructing
emission angles and kinetic energy the full four-momentum of the spectator
proton can be determined. 

The four-momentum of the actively scattered proton has been
obtained from its track information in either the forward or the central
detector. In the latter case the energy information was not retrieved. 

In the case that the actively scattered proton has been detected together with
its scattering
partner, the neutron (case (i)), we checked in addition,
whether the angular correlation for elastic kinematics is fulfilled.

That way we were able to reconstruct the full event, which includes also the
four-momentum of the neutron. In the case that the neutron was not measured,
the subsequent kinematic fit had one overconstraint in case (ii) and none in
case (iii). 
In the case that the neutron could be detected by a hit in the calorimeter
(composed of 1012 CsI(Na) crystals) of the central detector -- associated with
no hit in the preceding plastic scintillator barrel, the directional
information of the scattered neutron could be retrieved. Therefore, such
events, which correspond to case (i), have undergone a kinematic fit with two
overconstraints.

\subsection{Determination of the Beam Polarization}

In order not to distort the beam polarization, the magnetic field
of the solenoid in the central detector was switched off.
The measurements have been carried out with cycles of the beam polarization
"up", "down" and unpolarized (originating from the same polarized source), where
"up" and  "down" refers to a horizontal 
scattering plane. Runs with the conventional unpolarized source verified that
the beam originating from the polarized source indeed was unpolarized when
using it in its "unpolarized" mode.

The magnitude of the beam polarization has been determined
and monitored by $dp$ elastic scattering, which was measured in parallel by
detecting the scattered deuteron in the forward detector as well as the
associated scattered proton in the
central detector. In case of a transversally polarized deuteron beam the
dependence of the count rate $N(\Theta,\Phi)$ on the polar and azimuthal
scattering angles $\Theta$ and $\Phi$ is given by 

\begin{eqnarray}
N(\Theta,\Phi)\sim&&
1+\frac 3 2 P_zA_y(\Theta)cos\Phi+\frac 1 4 P_{zz}\\
 &&[A_{xx}(\Theta)(1-cos2\Phi)+A_{yy}(\Theta)(1+cos2\Phi)].~~~ \nonumber
\end{eqnarray} 

Here $P_z$ and $P_{zz}$ denote vector and tensor polarization of the deuteron
beam, whereas $A_y$, $A_{xx}$ and $A_{yy}$ are the respective vector and tensor
analyzing powers of the $dp$ scattering process.

With WASA covering the full azimuthal angular range, we may decompose
vector and tensor parts by fitting the observed azimuthal angular dependence
for specific polar angles by use of eq. (1). The absolute values for  
the vector and tensor components of the deuteron beam have
been obtained by fitting our results for the vector  
$A_y$ (Fig.~1, top) 
and tensor analyzing power 
$A_{yy}$ (Fig.~1, bottom) in absolute height 
to those obtained previously at ANL \cite{ANL} for $T_d$ = 2.0
GeV. Though this energy is somewhat below the one used here, the analyzing
powers in $pd$ scattering have been observed \cite{ANL,arvieux} to be only
weakly dependent on the beam energy. This is supported by very recent data
obtained at COSY-ANKE \cite{ANKE} at $T_d$ = 2.27 GeV. As a
result we obtain beam polarizations of $P_z = 0.67(2)$, $P_{zz}$ = 0.65(2)
for "up" and $P_z$ = -0.45(2), $P_{zz}$ = 0.17(2) for "down".

\begin{figure}
\centering
\includegraphics[width=0.99\columnwidth]{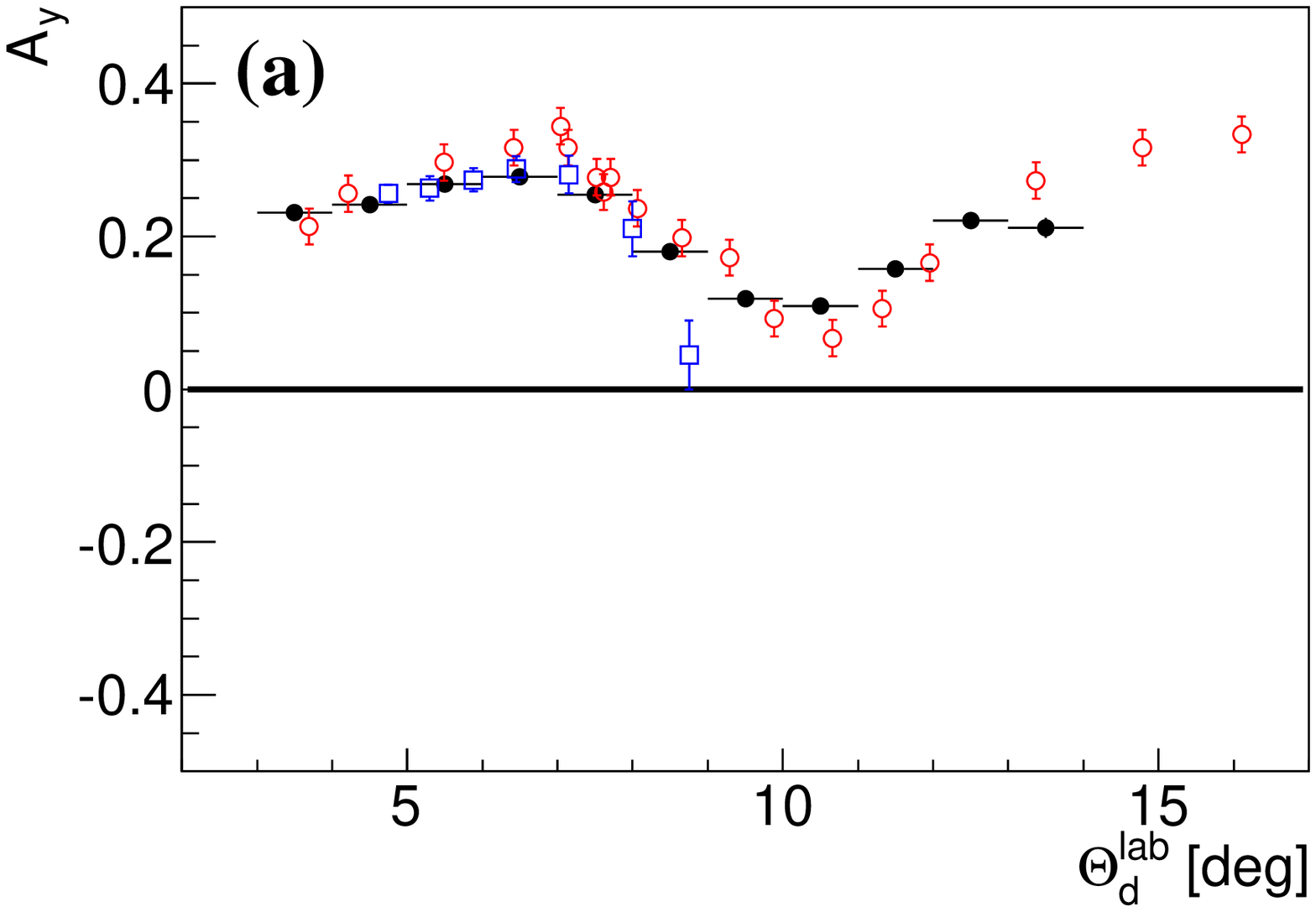}
\includegraphics[width=0.99\columnwidth]{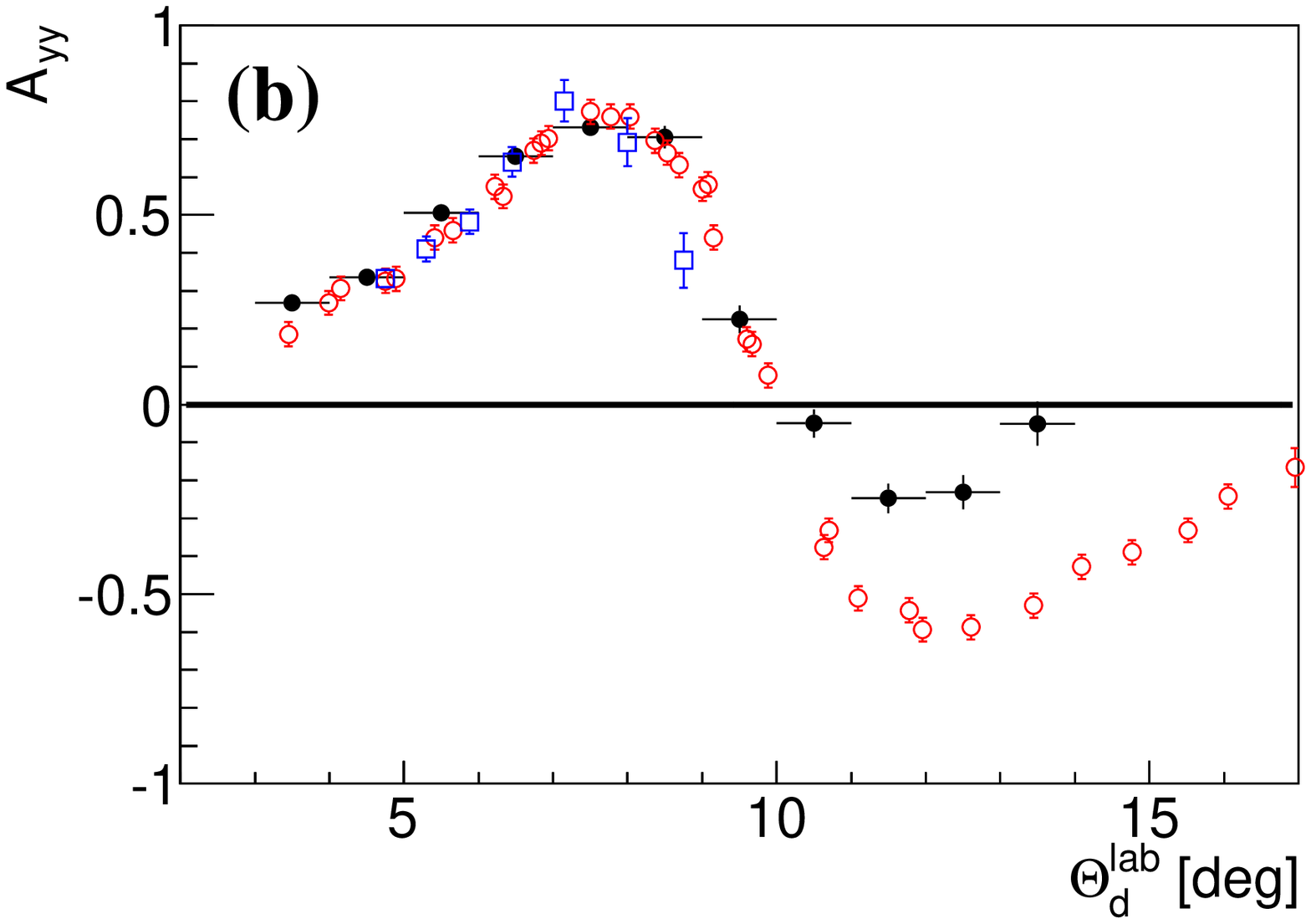}
\caption{\small (Color online) Angular distributions of vector ($A_y$, top)
  and tensor ($A_{yy}$, bottom) 
  analyzing powers in $dp$ scattering at $T_d$ = 2.27 GeV. The filled circles
  denote results from this  work, whereas open symbols denote results
  previously obtained at Argonne National Lab at $T_d$ = 2.0 GeV \cite{ANL}
  (open circles) and COSY-ANKE at $T_d$ = 2.27 GeV \cite{ANKE}(open
  squares). The shown error bars denote statistical uncertainties.
}
\label{fig1}
\end{figure}

\begin{figure}
\centering
\includegraphics[width=0.99\columnwidth]{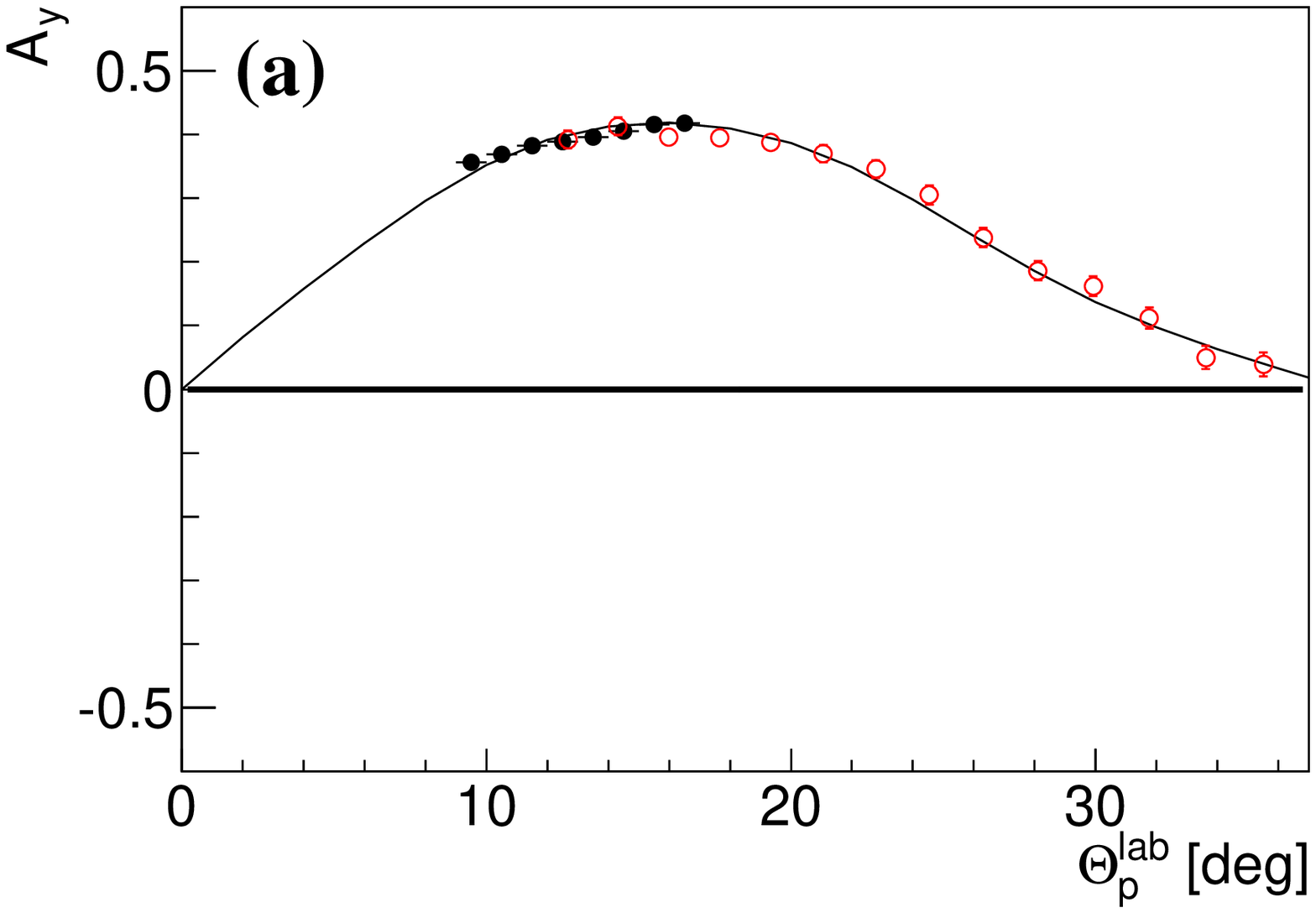}
\includegraphics[width=0.99\columnwidth]{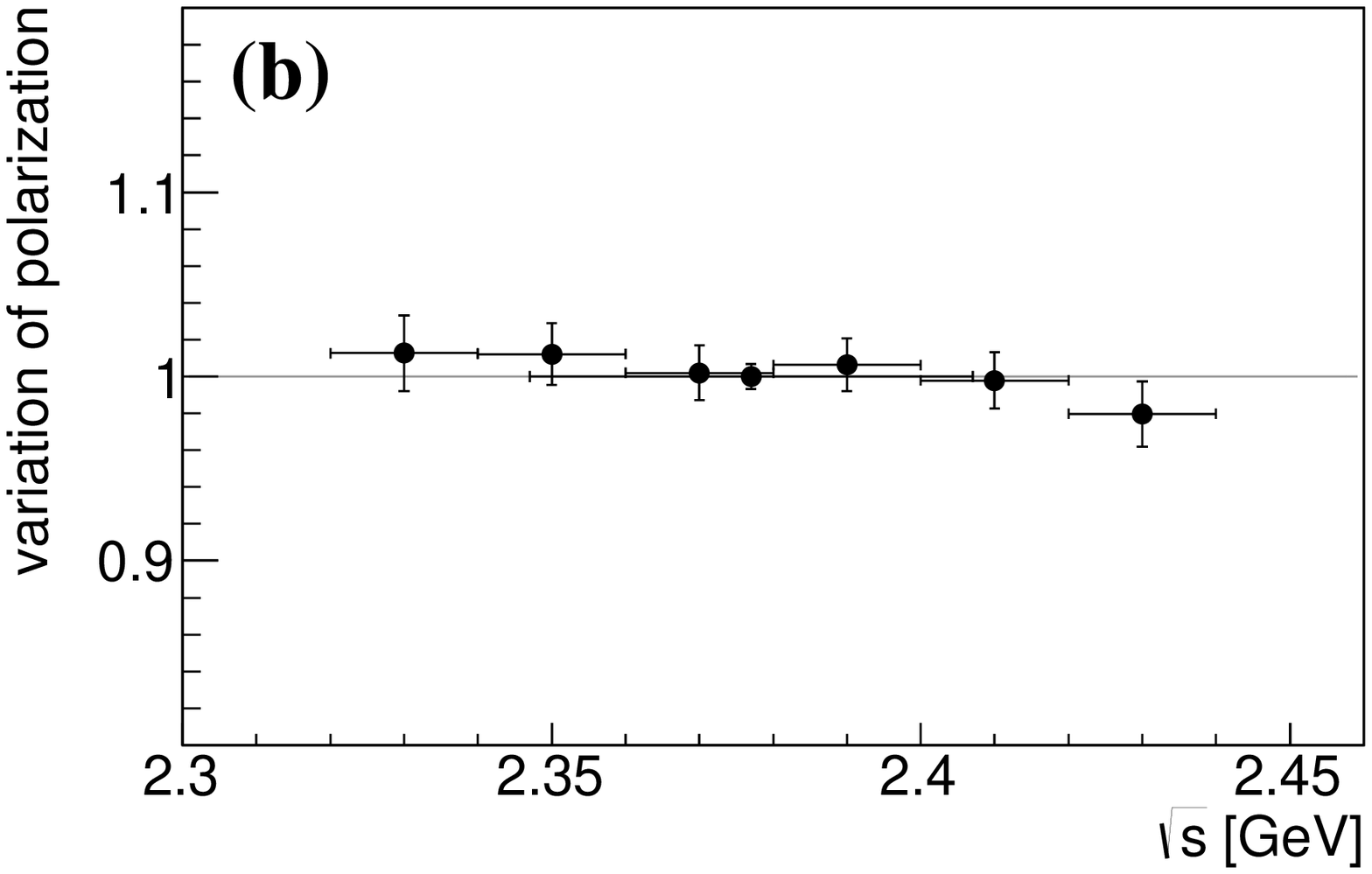}
\caption{\small (Color online) Top: 
  Vector analyzing power $A_y$ in 
  $pp$ scattering at $T_p$ = 1.135 GeV. Filled circles denote results
  from this work, whereas open symbols denote results previously obtained with
  EDDA at $T_p $ = 1.1316 GeV \cite{EDDA}. The solid line represents the SAID SP07 phaseshift solution
  \cite{Arndt07}. Bottom: Variation of the proton polarization over the
  measured energy interval in relation to the SAID SP07 phase shift
  solution. The datum at 2.377 GeV derives from the figure shown at the top
  and denotes the average over the full energy range covered in this experiment.
}
\label{fig2}
\end{figure}

\subsection{Checks of Quasifree Scattering}

The vector polarization of the beam for quasifree scattering has been checked
by quasifree $pp$ scattering, which also was measured in parallel by detecting
one of the protons in the forward detector and the other one in the central
detector -- and, in addition, checking their angular correlation for elastic
kinematics. If the proton in the beam deuteron is at rest, then its momentum
corresponds to just half of the beam momentum. Note that in the energy region of
interest here, the $pp$ analyzing power does not exhibit any significant
energy dependence, hence  we do not need to correct for the energy smearing
due to the Fermi motion of the proton.   
Fig.~2 shows our results from the quasifree $pp$ scattering for the analyzing
power. On top the angular distribution is shown in comparison  with the EDDA
measurements \cite{EDDA} of free $pp$ 
scattering. The solid curve gives the SAID phase shift solution SP07
\cite{Arndt07}, which again is based in this energy range on the EDDA data. 

In order to check, whether the proton polarization depends on the effective
energy of the quasifree incident protons, we reconstruct the effective
center-of-mass energy $\sqrt s$ for each event. That way we obtain angular
distributions sorted in six $\sqrt s$ bins, which we compare to SAID SP07 phase
shift predictions for these energies. The ratio of our $pp$ data to the SAID 
prediction at each of these energies is plotted in Fig.~2, bottom. Within
uncertainties no deviation from unity is observed, {\it i.e.}, the proton
polarization can be considered to be constant within 1 $\%$ over the energy
interval covered by this experiment. 

\begin{figure} 
\centering
\includegraphics[width=0.99\columnwidth]{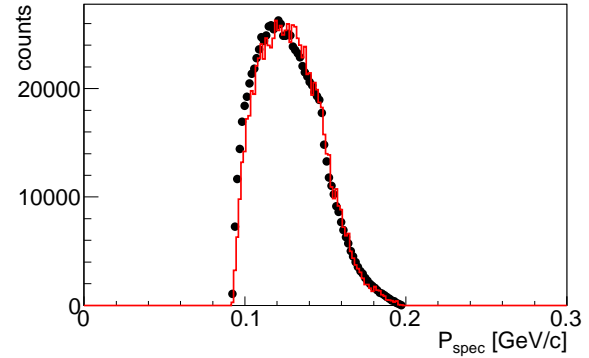}
\caption{\small (Color online)
  Distribution of the spectator proton momenta (in the deuteron rest frame) in
  the $dp \to pn +  
  p_{spectator}$ reaction within the acceptance of the WASA detector. Data are
  given by the full circles. The solid line shows 
  the expected distribution for the quasifree process based on the CD Bonn 
  potential \cite{mach} deuteron wavefunction. 
  For the data analysis only events with $p_{spectator} <$ 0.16 GeV/c
  have been used.
}
\label{fig3}
\end{figure}

The momentum distribution of the observed spectator proton in the elastic $np$
scattering process is plotted in Fig.~3 in the deuteron rest frame, where it
is compared with Monte Carlo simulations of the proton momentum distribution
in the deuteron filtered by the acceptance of the WASA detector. 
In these simulations the CD Bonn potential \cite{mach} deuteron wavefunction
has been used. Due to the beam-pipe
ejectiles can only be detected in the forward detector for lab angles larger
than three degrees. In order to assure a quasi-free process we omit events
with  spectator momenta larger than 0.16 GeV/c ( in the deuteron rest system)
from the subsequent analysis -- as done in previous work \cite{MB,isofus,TS}.

\section{Results and Discussion}

\subsection{ $A_y$ Angular Distributions}

Since we have measurements with spin "up", "down" and unpolarized, the vector
analyzing power for $np$ scattering can be derived in three different ways by
using each two of 
the three spin situations. All three methods should give the same
results. Differences in the results may be taken as a measure of systematic
uncertainties, which are added quadratically to the statistical ones to give
the total uncertainties.

In Fig.~4 we show the results for
$A_y$, if we either combine measurements with spin "up" and unpolarized (open
circles) or with spin "down" and unpolarized (solid squares) for extracting
$A_y$. The values for the combination "up'' and "down'' are just in between
(solid circles). For this plot the data are used without  accounting for the
spectator momentum, {\it i.e.} without selection according
to the $np$ center-of-mass energy. Thus this data set corresponds to the
weighted average over the covered interval of $\sqrt s$.

Due to the Fermi motion of the nucleons bound in the beam deuteron, the
measurement of the quasi-free $np$ scattering process covers a range of
energies in the $np$ system. Meaningful statistics could be collected for
the range of $np$ center-of-mass energies 2.37 $< \sqrt s <$ 2.40 GeV
corresponding to $T_n$ = 1.11 - 1.20 GeV. 

By taking into account the measured spectator four-momentum we may construct
the effective $\sqrt s$ for each event. That way we obtain angular
distributions sorted in $\sqrt s$ bins. In order to have sufficient count-rate
statistics we restricted this procedure to the data sets in the angular region
$31^\circ < \Theta_n^{cm} <
  129^\circ$ and divided the available energy range into six bins, which are
  shown in Fig.~5 together with the SAID SP07 solution (solid 
lines) and the new solution -- see next section -- which contains a resonance pole (dashed and dotted
lines). We also include in Fig.~5 previously obtained angular distributions at
$\sqrt s$ = 2.360 and 2.440 GeV ($T_n$ = 1.095 and 1.27 GeV)
\cite{ball,les,mak,dieb}, which are closest to the resonance region covered
here. Note, that these angular distributions being below and above the
resonance region exhibit a significantly different angular behavior -- in
particular at medium angles.

\begin{figure}
\centering
\includegraphics[width=0.99\columnwidth]{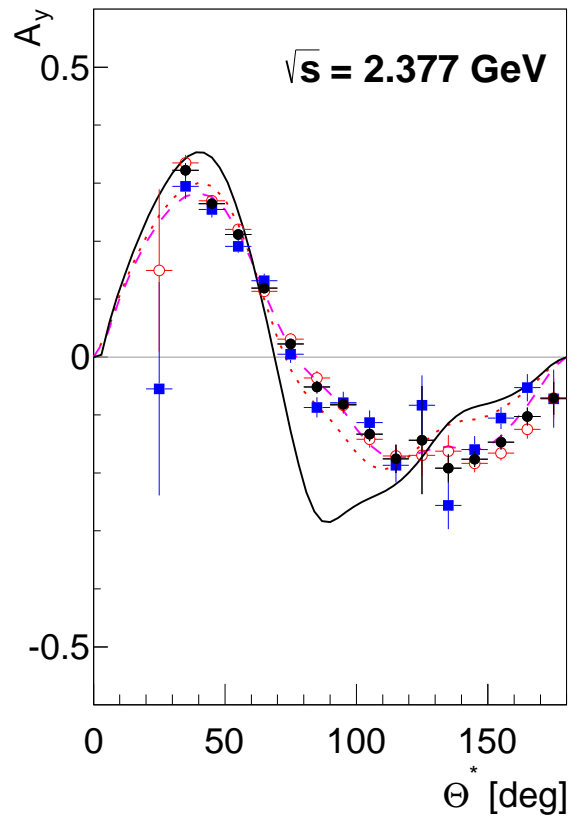}
\caption{\small (Color online) Angular distributions of the $np$ analyzing power
  without consideration of the spectator momentum, {\it i.e.} 
  without classifying 
  the collected $pn$ scattering events according to their effective total
  center-of-mass energy $\sqrt s$. That way the data set corresponds to the
  weighted average over the measured interval representing effectively the
  range $\sqrt s $ = 2.377$\pm$0.03 GeV 
  (corresponding to 1.075 GeV $\leq T_n \leq$ 1.195 GeV). 
  Open circles and solid squares denote the $A_y$
  extraction by the combination "up" with "unpolarized" and "down" with
  "unpolarized", 
  respectively. The solid circles give the statistically weighted average over
  both methods. The error bars denote statistical uncertainties. 
  The solid line 
  represents the SAID SP07 phase shift solution \cite{Arndt07}, whereas the
  dashed (dotted) line gives the result of the new weighted (unweighted) SAID
  partial-wave solution, see text.  
}
\label{fig4}
\end{figure}

\begin{figure*}
\centering
\includegraphics[width=0.89\columnwidth]{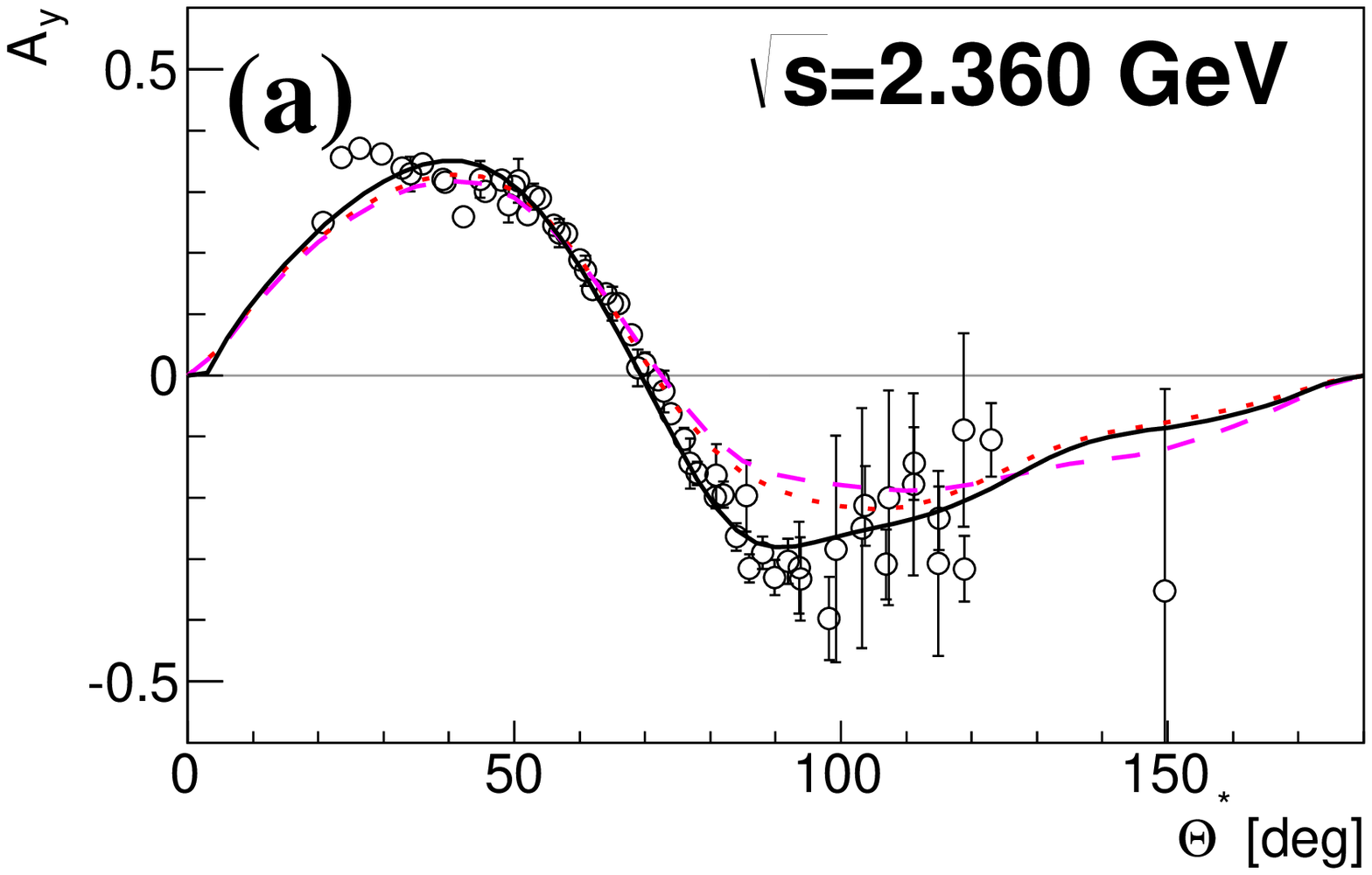}
\includegraphics[width=0.89\columnwidth]{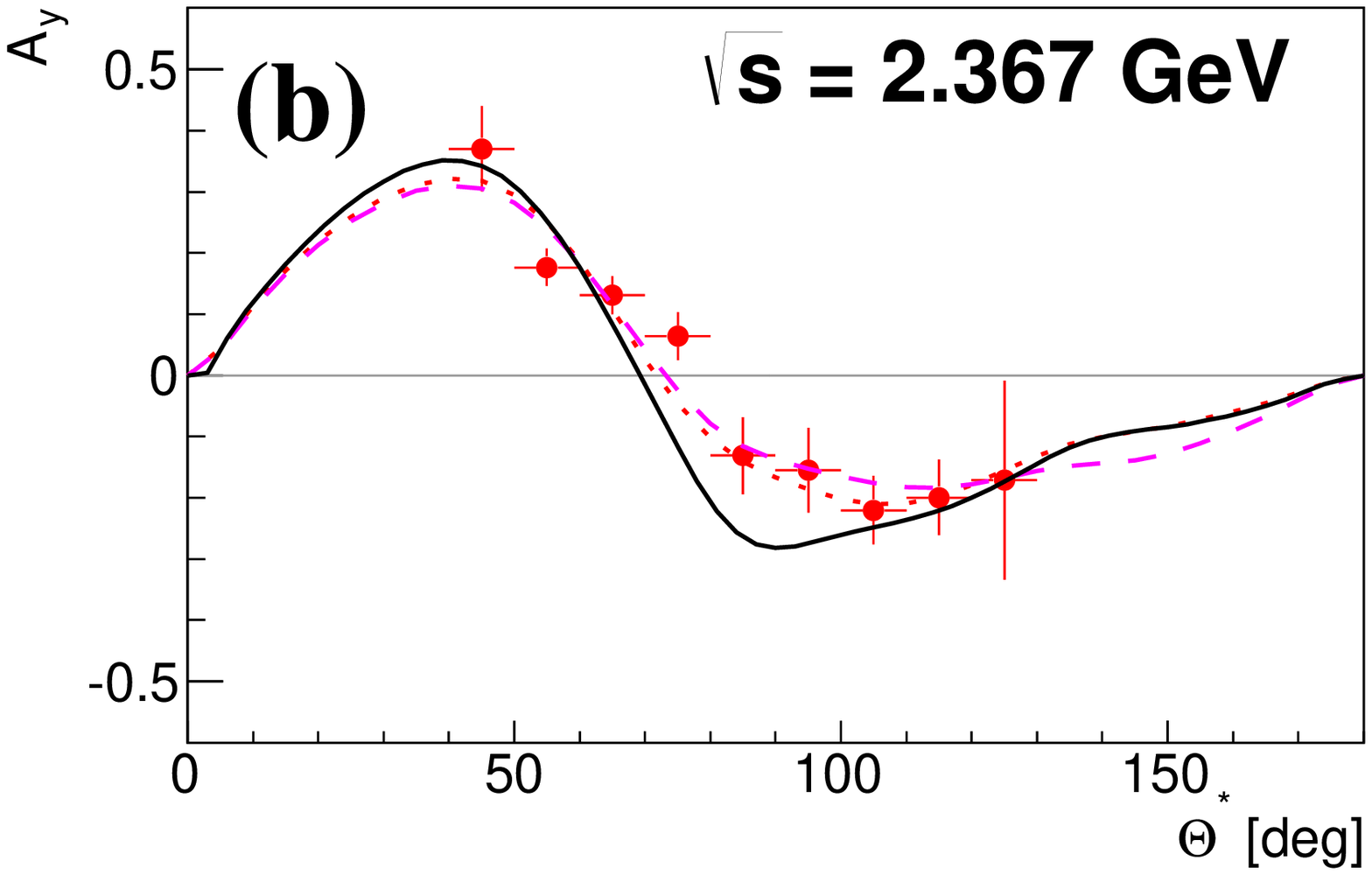}
\includegraphics[width=0.89\columnwidth]{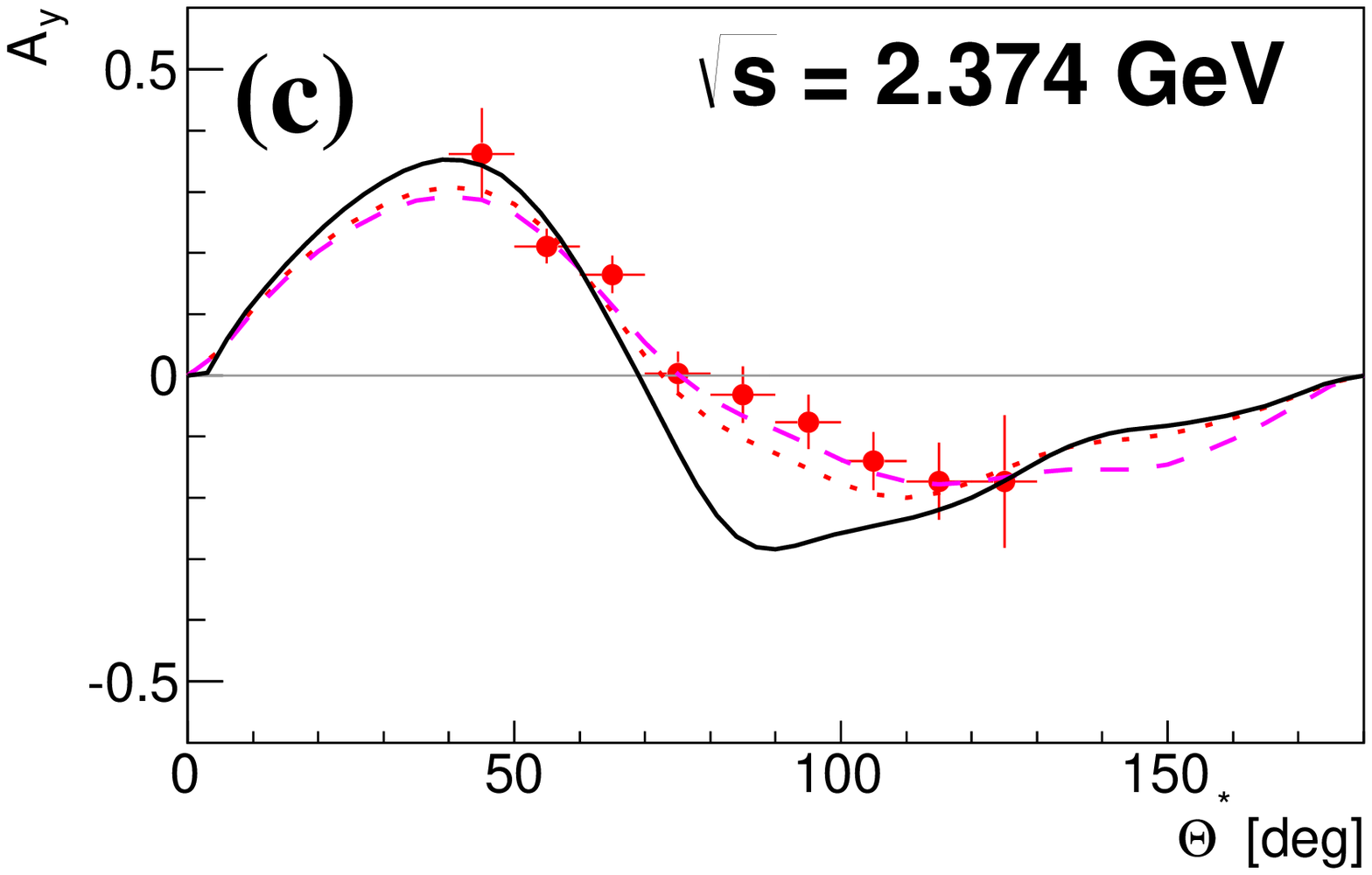}
\includegraphics[width=0.89\columnwidth]{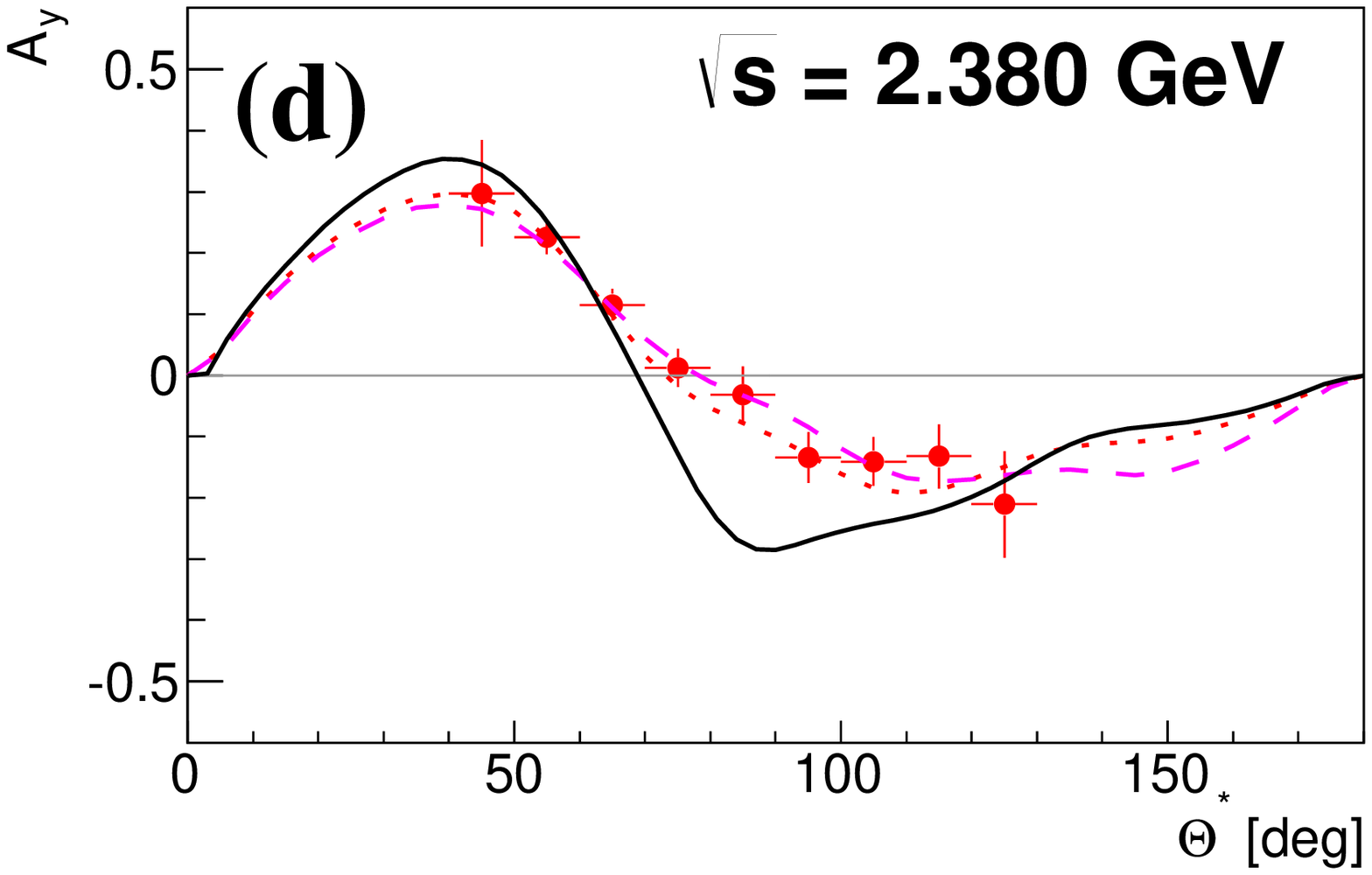}
\includegraphics[width=0.89\columnwidth]{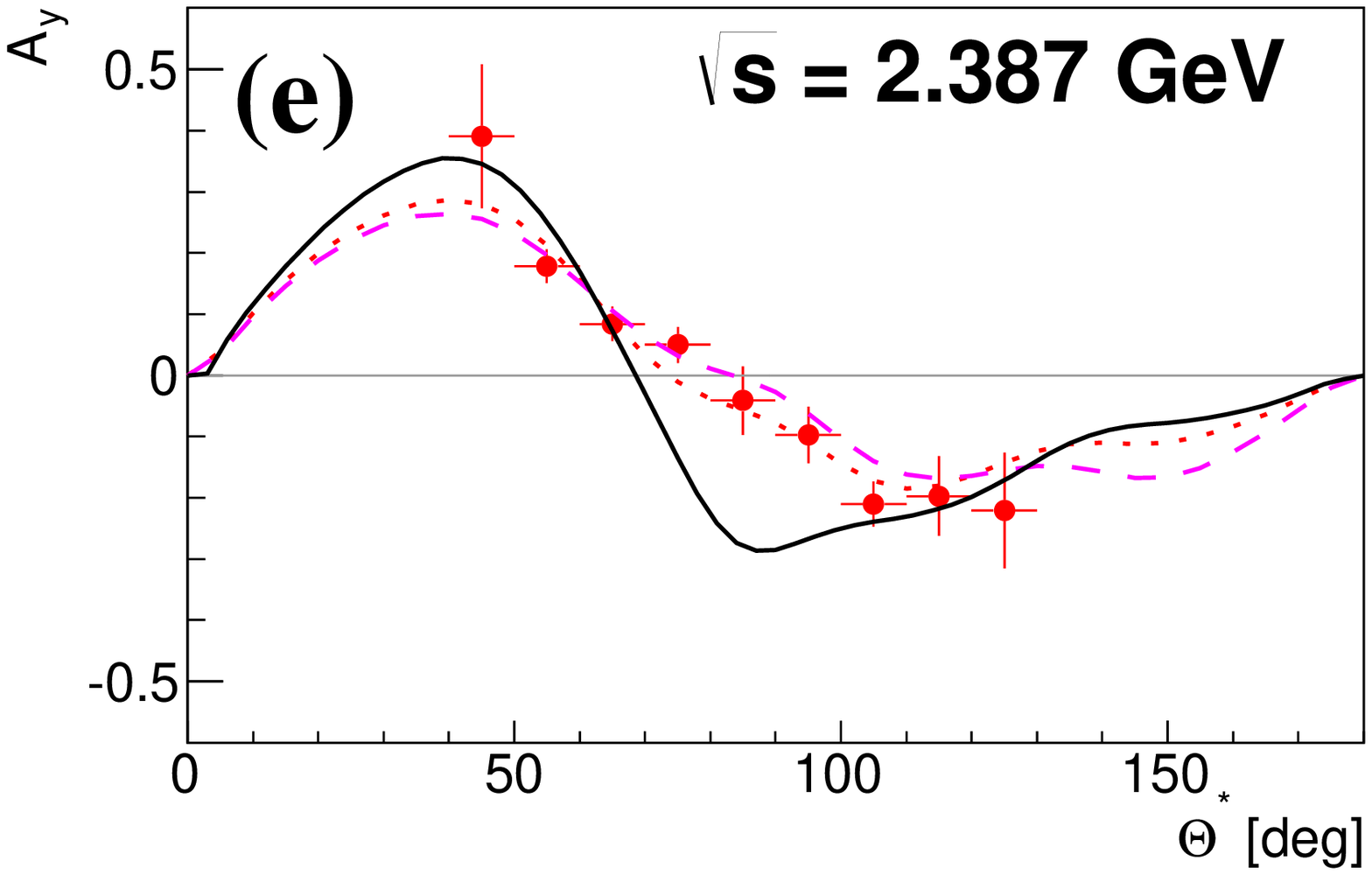}
\includegraphics[width=0.89\columnwidth]{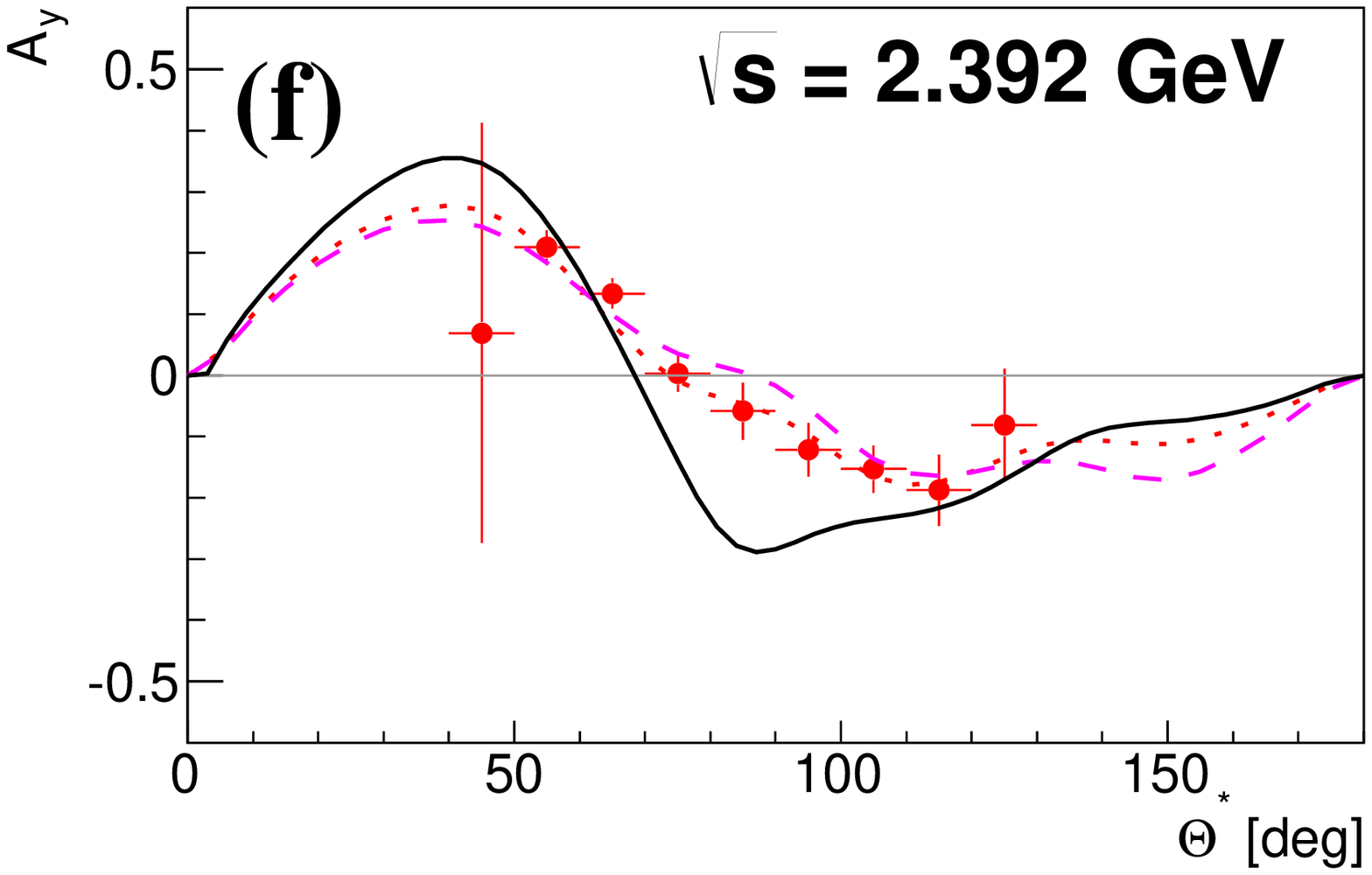}
\includegraphics[width=0.89\columnwidth]{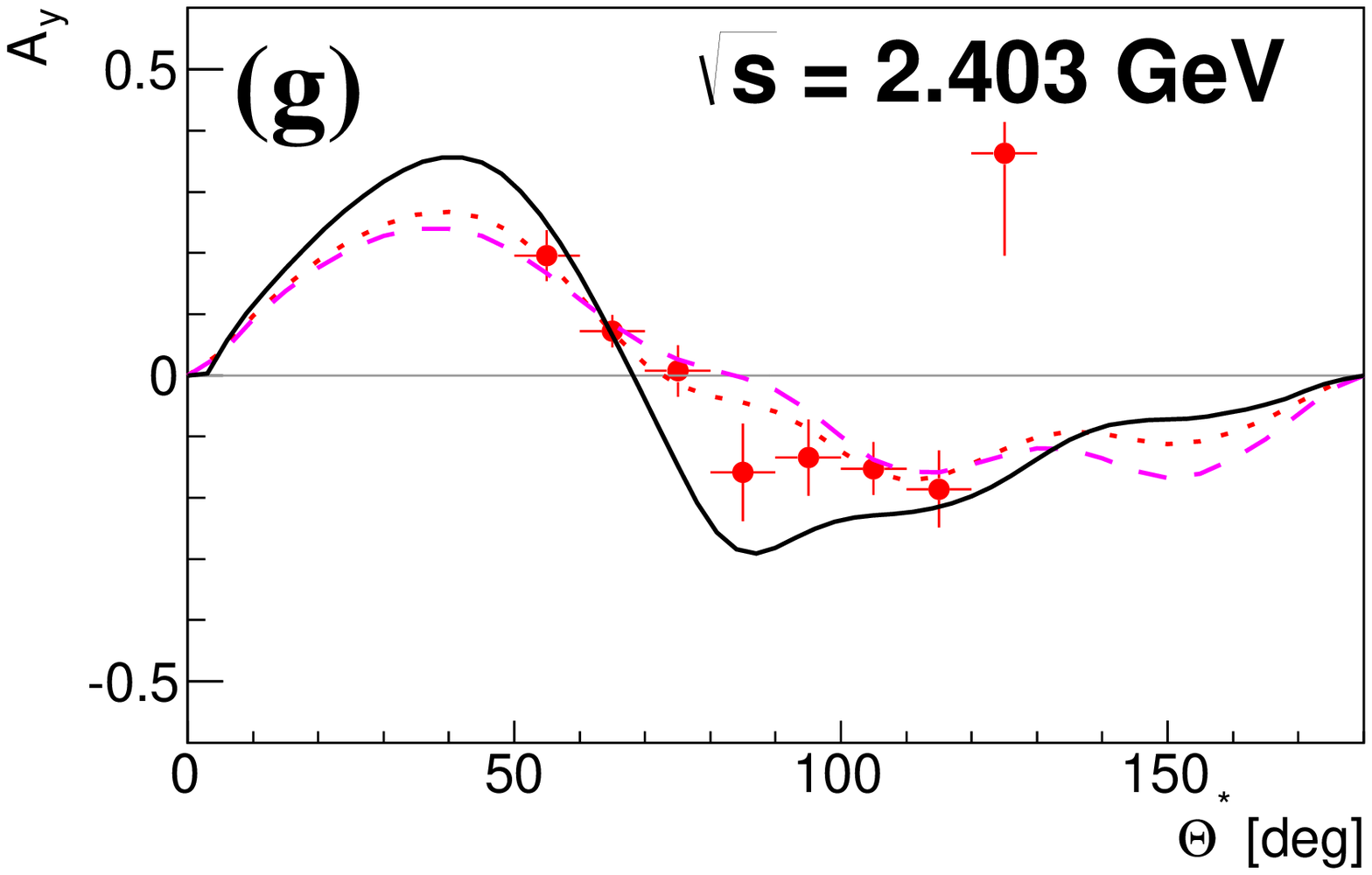}
\includegraphics[width=0.89\columnwidth]{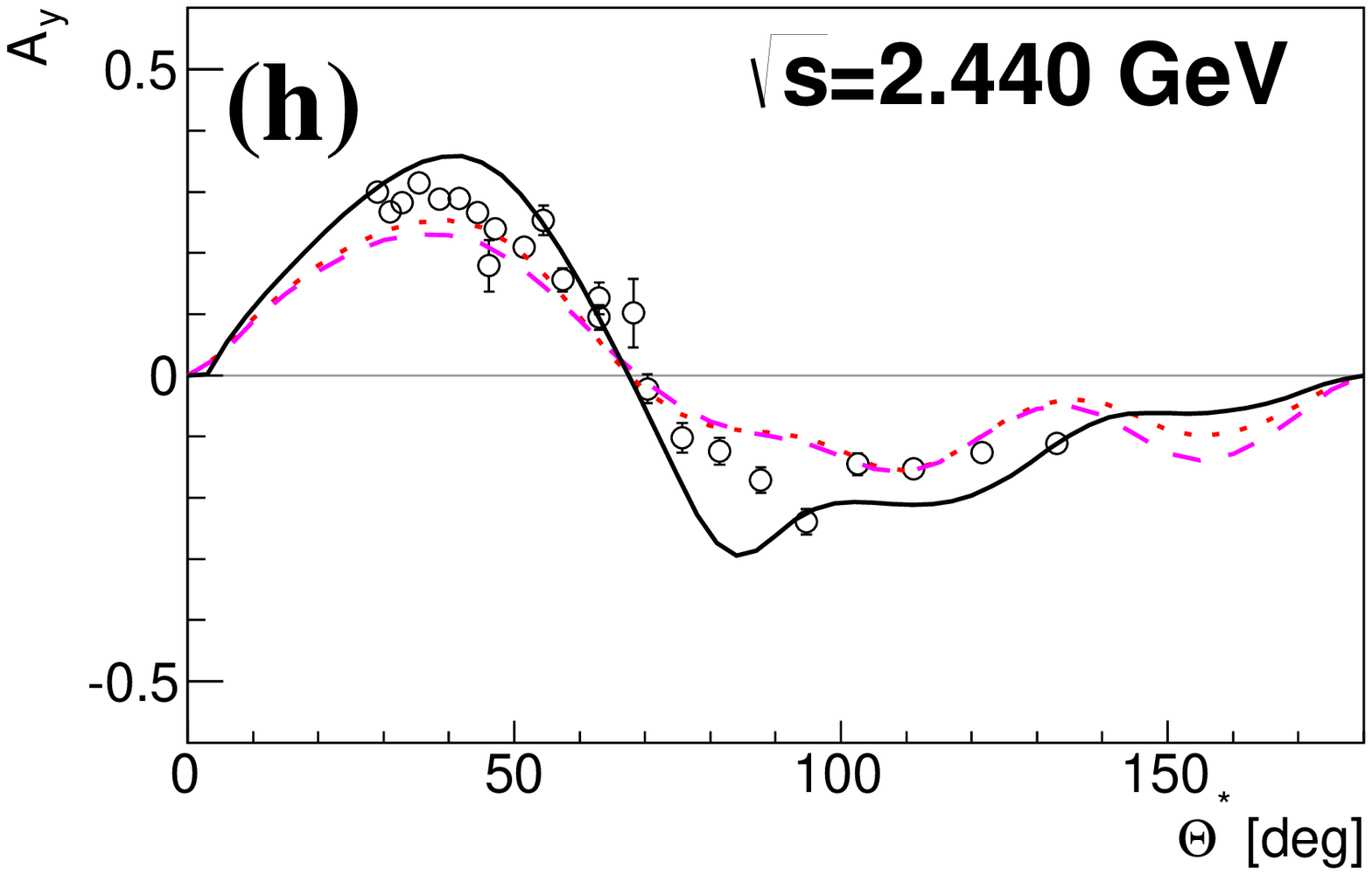}
\caption{\small (Color online) $A_y$ angular distributions for  $\sqrt s$ =
  2.360, 2.367, 2.374, 2.380, 2.387, 2.392, 2.403
  and 2.440 GeV corresponding to $T_n$ = 1.095, 1.108, 1.125, 1.139, 1.156,
  1.171, 1.197 and 1.27 GeV. The full symbols 
  denote results from this work taking into account the spectator
  four-momentum information. The error-bars on the full symbols include both
  statistical and 
  systematic uncertainties. The open symbols denote previous work
  \cite{ball,les,mak,dieb}. The solid lines represent the SAID SP07 phase
  shift solution \cite{Arndt07}, whereas the dashed (dotted) lines give the
  result of the new weighted (unweighted) SAID partial-wave solution, see text.
}
\label{fig5}
\end{figure*}

In Fig.~6 we show as an example the energy dependence of $A_y$ at
$\Theta_n^{cm} = 70^\circ$ --- see also Fig.~4 in Ref. \cite{prl2014}, where
the energy dependence  at $\Theta_n^{cm} = 83^\circ$ is depicted. The trend of
the new data in the resonance region 
deviates clearly from that exhibited by the data from previous experiments
below and above this region
\cite{ball,les,mak,new,ball1,arn,mcn,gla,bar}. The new partial-wave solutions
(see below) connect all data within their uncertainties, whereas the SP07
solution obviously fails in the resonance region.  

\begin{figure}
\centering
\includegraphics[width=0.99\columnwidth]{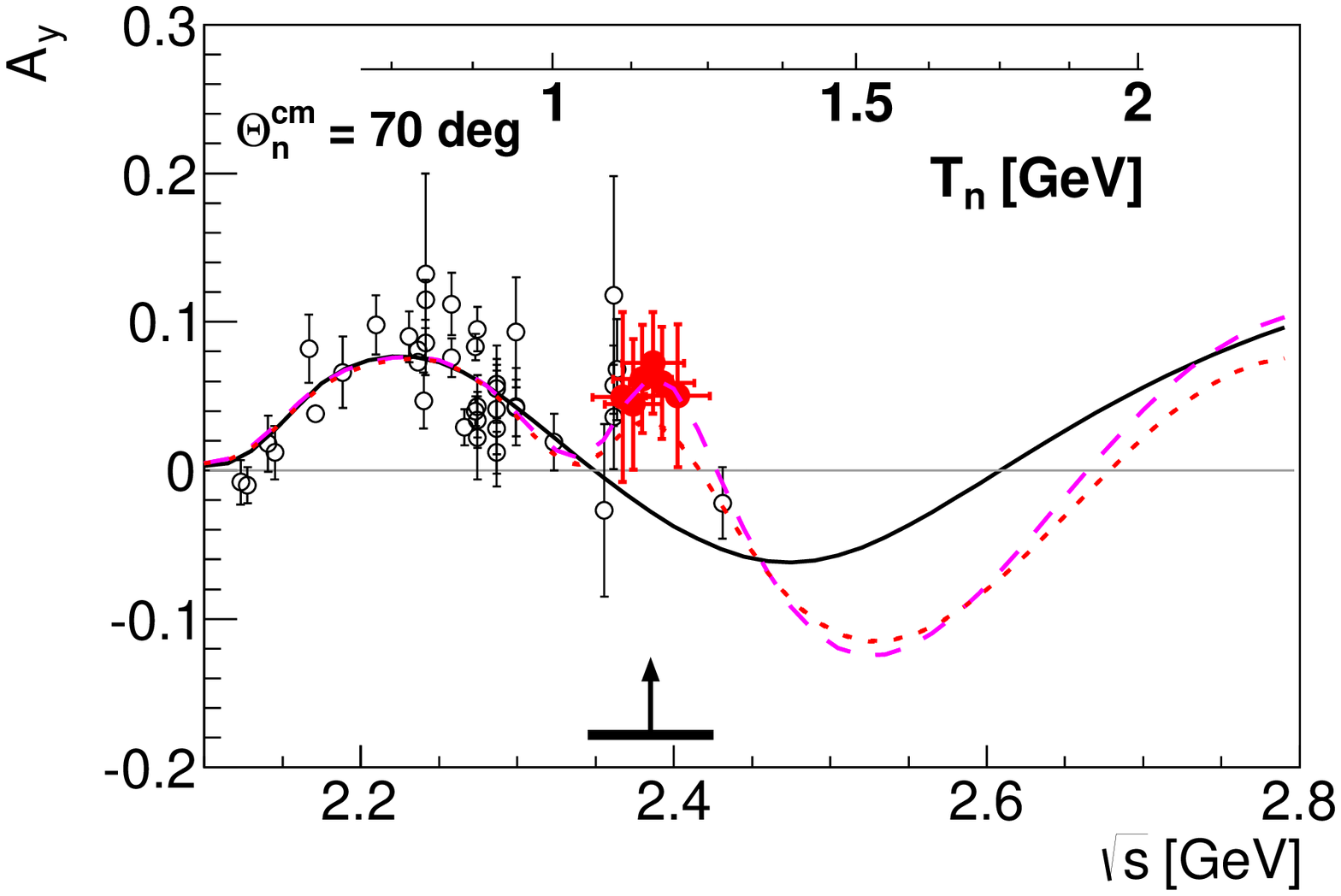}
\caption{\small (Color online) Energy dependence of the $np$ analyzing power
  at $\Theta_n^{cm} = 70^\circ$. The solid symbols denote the results of this
  work, the open symbols denote those from previous work
  \cite{les,mak,ball,new,ball1,arn,mcn,gla,bar}. For the
  meaning of the curves see Fig.~4. Vertical arrow and horizontal bar
  indicate pole and width of the resonance.
}
\label{fig3}
\end{figure}

The decomposition of the $np$-scattering observables into partial wave
amplitudes is given in Ref. \cite{ArndtRoper}. Accordingly we have for the
analyzing power:

\begin {equation}
 d\sigma / d\Omega * A_y \sim Im{(H_3 + H_5) H_4^*}
\end {equation}

with $H_i$ containing sums over partial wave amplitudes with total angular
momenta $j_0 = j = L$, $j_- = L - 1$ and $j_+ = L + 1$. $H_3$ contains terms
being proportional either to the Legendre polynomials $P_j$ or to the associated
ones $P_j^1$. In $H_5$ there are terms only proportional to $P_j$ and in $H_4$
only proportional to $P_j^1$. In particular, the structure of $H_4$ for $j =
3$ is as follows:
\begin {equation}
H_4(j = 3) \sim [4(T_{L=4} - 3 T_{L=2}) + \sqrt{12} T_{L=3}] P_3^1,  
\end {equation}
where the T-matrix elements contain the complex phase shifts. We see that a
resonance effect in $^3D_3$ and $^3G_3$ enters with opposite sign and is
proportional to $P_3^1$ in both cases. Hence the resonance effect vanishes at
the zeros of $P_3^1$, which is the case at $\Theta = 63.4^\circ$ and
116.6$^\circ$. At these angles the predictions with and without resonance in
$^3D_3$ or $^3G_3$ should cross each other. If in the SP07 solution and in the
new solution the 
non-resonant contributions are comparable, then the angular distributions
calculated with these solutions should cross at these angles. Fig.~5 
demonstrates that this is the case in good approximation.

By observing the  maximum deviations from the SP07 solution in the angular
region around 90$^\circ$ as well as the minimum deviations around 63$^\circ$ and
117$^\circ$ -- coupled with a sign change thereafter -- both data and new SAID
solutions exhibit the characteristic features of the $P_3^1$ function and thus
uniquely point to the signature of a $J^P = 3^+$ resonance in the elastic $np$
scattering.  

The horizontal bars on the data points in Fig.~6 (and also in Fig.~4 of
Ref.~\cite{prl2014}) include both the range of the $\sqrt s$ bins  and the
uncertainties in the determination of the $\sqrt s$ values reconstructed for
each event. Since we deal here only with 0 - 2 overconstraints in the
kinematic fits, the $\sqrt s$ determination is less precise than, {\it
  e.g.}, in the $pd \to d\pi^0\pi^0 + p_{spectator}$ reaction, where we have
three overconstraints in the case that the proton spectator is not detected.


\subsection{Partial-Wave Analysis}

The new $A_y$ data have been included in the SAID database and the
phenomenological approach used in 
generating the $NN$ partial-wave solution, SP07~\cite{Arndt07},
has been retained.
Here we are simply considering whether the existing form is
capable of describing the new $A_y$ measurements. One advantage
of this approach is that the employed Chew-Mandelstam K-matrix 
can produce a pole in the complex energy plane without the 
explicit inclusion of a K-matrix pole in the fit form. Neither
the existence of a pole nor the effected partial waves are 
predetermined.


The energy-dependent fits use a product S-matrix approach 
      as described in detail in Ref. \cite{Arndt87}
with $S_x$ being an 'exchange' part, including the one-pion-exchange
piece, plus smooth phenomenological terms, and $S_p$, a
'production' part. The full S matrix is
\begin{equation}
S = S_x^{1/2} \; S_p \; S_x^{1/2} = 1 \; + \; 2 i T,
\end{equation}
where
\begin{equation}
T = T_x + S_x^{1/2}\; T_p \; S_x^{1/2} .
\end{equation}
For spin-uncoupled waves, the production T-matrix is
parameterized using a Chew-Mandelstam K-matrix, as is also
used in the GW $\pi N$~\cite{pin} and $K N$~\cite{kn} analyses, with
\begin{equation}
T_p = \rho^{1/2} K_p (1-CK_p)^{-1} \rho^{1/2},
\end{equation}
where $\rho$ is a phase space factor, $K_p$ is a real symmetric
matrix coupling the $NN$ and an inelastic channel, and $C$ is
a Chew-Mandelstam matrix. For isovector waves, the inelastic channel is
identified as $N \Delta$; in the isoscalar case, this inelastic channel is
generic. 
For spin-uncoupled waves, the matrices
are $2\times 2$; for coupled waves, the matrices are $3\times 3$,
as described in Ref. \cite{Arndt87}.
The global energy-dependent fit includes $pp$ data from 
threshold up to a lab kinetic energy of 3 GeV, and $np$ data from
threshold up to 2 GeV. Since above 1.3 GeV the amount of $np$ data is sparse,
the fit is considered to be valid only up to 1.3 GeV for the $np$ case.
Single-energy (narrow energy bin) fits
are also carried out, with constraints on the energy-dependence over
a particular energy bin fixed to the underlying global analysis. 

The new $A_y$ data are angular distributions at $T_{\rm Lab}$ values
of 1.108, 1.125, 1.135, 1.139, 1.156, 1.171, and 1.197 GeV. 
Starting from the functional form of the current SP07 fit, and only varying
the associated free parameters, a $\chi^2$/datum of 1.8 was
found for all angular distributions apart from the one at
1.135 GeV. This is fairly consistent with the overall $\chi^2$/datum
given by the global fit of $np$ to 2 GeV. However, the set
at 1.135 GeV contributes a $\chi^2$/datum of about 25, has
better statistics and a wider angular coverage. 

The fit form was scanned to find partial waves for which
an added term in the K-matrix expansion produced the most
efficient reduction in $\chi^2$. Adding parameters and
re-fitting resulted in a rapid variation of the coupled
$^3D_3$ and $^3G_3$ waves in the vicinity of the problematic
1.135 GeV data set.  

Some weighting seemed necessary in this fit, as only a few
angular points from the full set were determining the 
altered energy dependence.  The
fit was repeated with different weightings (with a factor of 2 or 4) for the
new $A_y$ data.
Initially, the full set of 
energies was weighted equally. However, it was found that
just weighting the 1.135 GeV angular distribution improved the
fit to the new analyzing power data at all energies. The
results reported here thus considered only the weighting at
this single energy.
As we have seen in fits to other reactions, heavily weighting
new and precise polarization observables inevitably degrades
the fit to older data. Therefore, as a test, 
the parameterization producing
a pole was re-fitted to the full database with no weighting. This
gave, as expected, a worse fit to the 1.135 GeV angular distribution
but did not change the shape qualitatively.



In Figs.~4 - 6  and 9 -11 we plot the
SP07 prediction (not including the new data), a weighted fit (errors
decreased by a factor of 4), and an unweighted fit including the
new data and using the fit form having added parameters.

Resulting changes in the $^3D_3$-$^3G_3$ coupled waves have been
displayed already in Ref.~\cite{prl2014}, Fig.~3. 
Note that the single-energy solutions obtained previously for energies up 1.1
GeV fit better to the new partial-wave solution than to SP07. 

In the new solution the $^3D_3$ wave obtained a typical
resonance shape, whereas the $^3G_3$ wave changed less dramatically.
In Fig.~7 the Argand diagrams of the new partial-wave solution are shown for
$^3D_3$ and $^3G_3$ partial waves as well as for their mixing amplitude
$\epsilon_3$. In the Argand diagram the $^3D_3$ partial wave exhibits an
abrupt change at  
the pion production threshold ($\sqrt s$ = 2.02 GeV), when absorption sets in,
followed by a pronounced looping in the $d^*$ energy region, before it
enters the region of 
the conventional $t$-channel $\Delta\Delta$ process \cite{mb,MB,isofus,TS} at
the highest energies. In the Argand diagrams of $^3G_3$ and $\epsilon_3$ also
a looping is observed in the $d^*$ region, though much less pronounced.

\begin{figure}
\centering
\includegraphics[width=0.99\columnwidth]{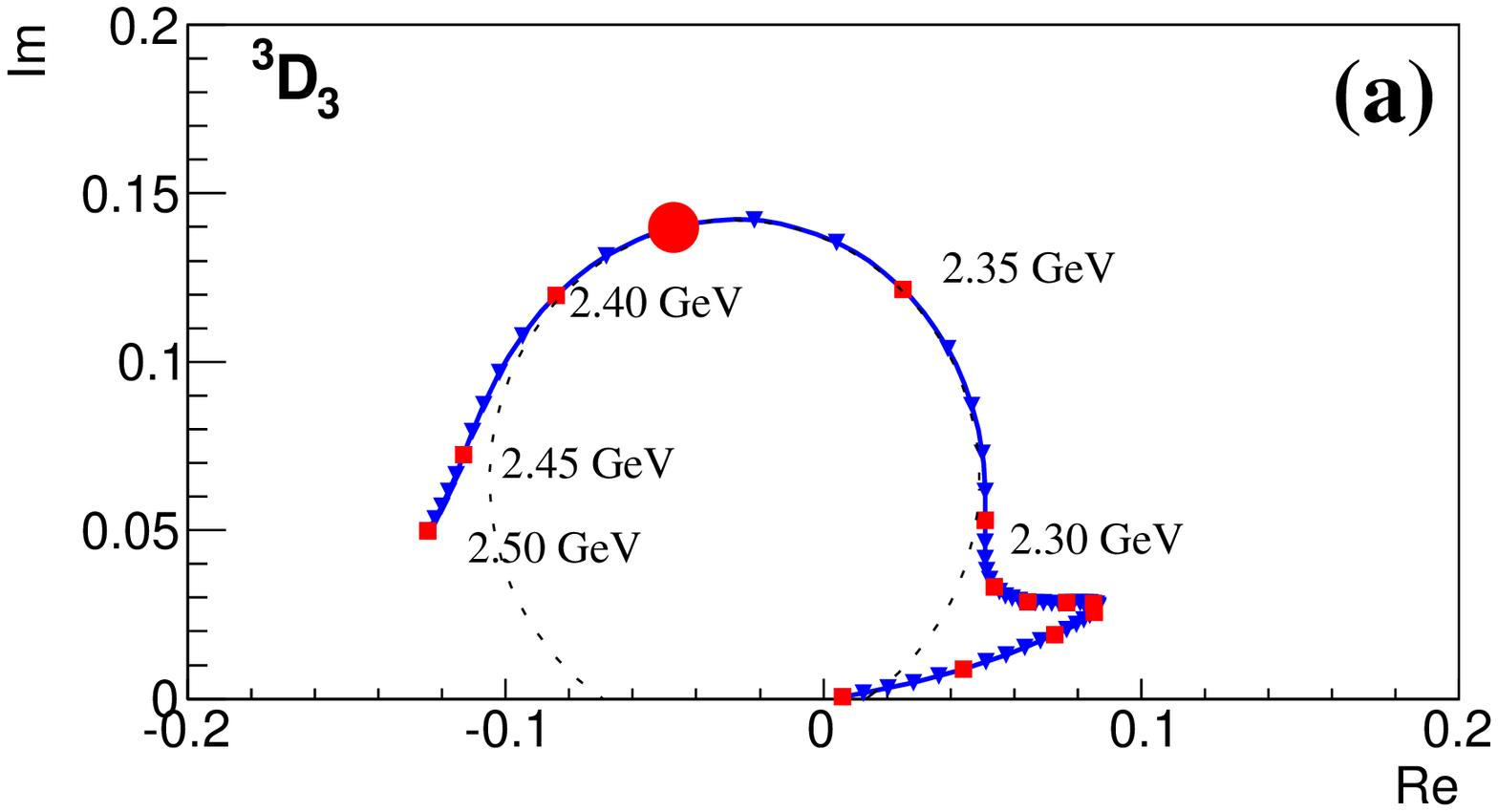}
\includegraphics[width=0.69\columnwidth]{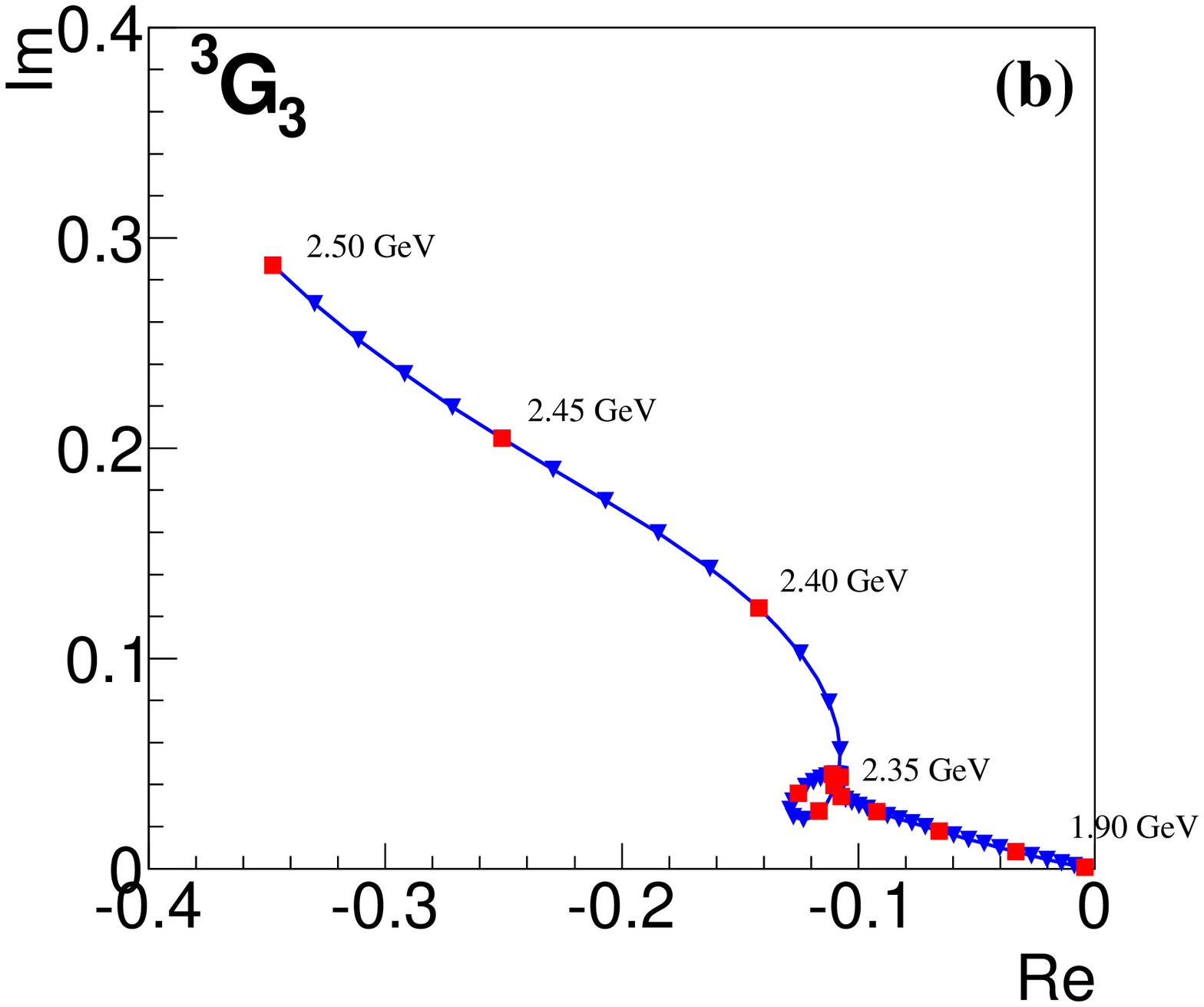}
\includegraphics[width=0.69\columnwidth]{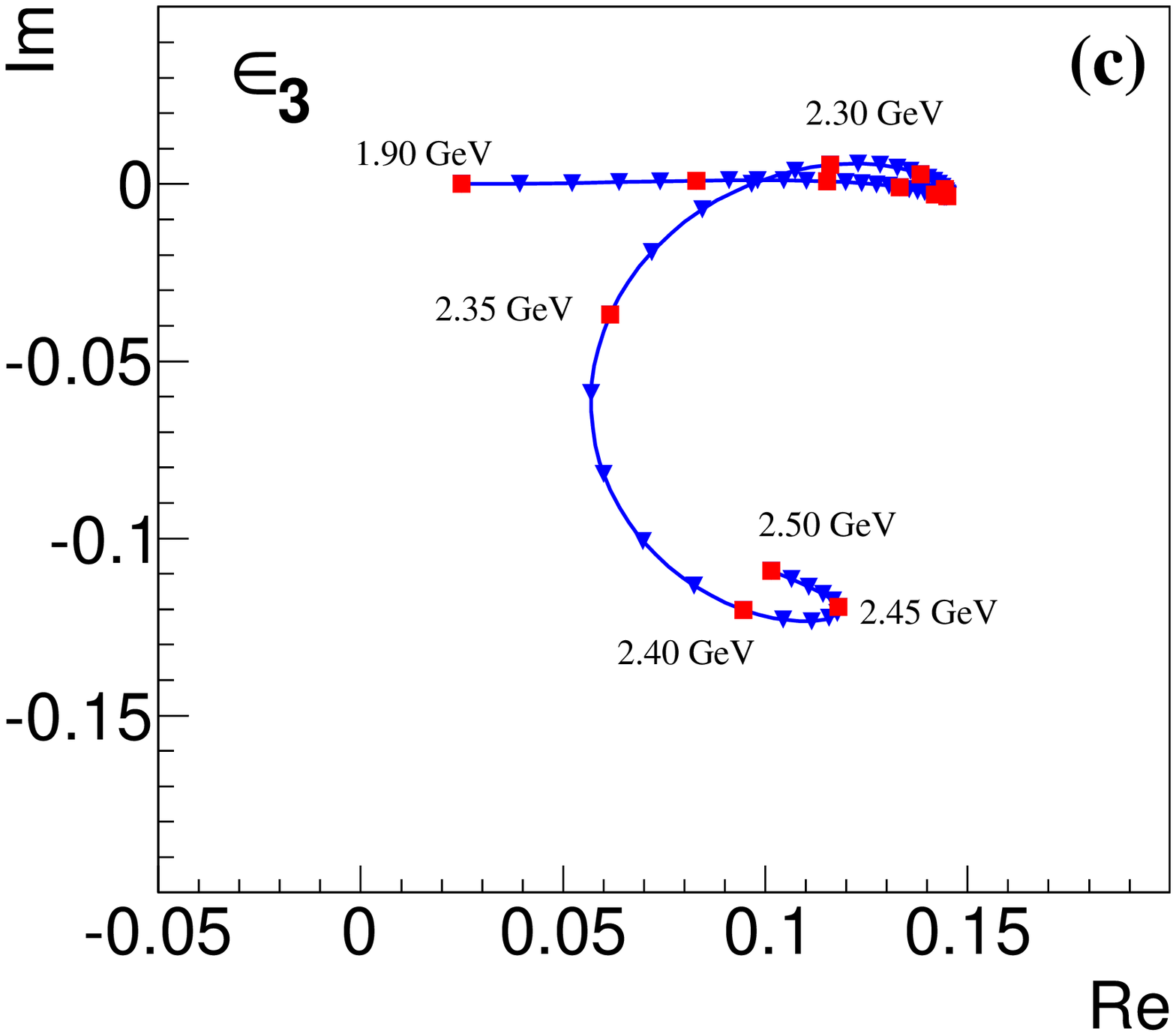}
\caption{\small (Color online) Argand diagrams of $^3D_3$ (top) and $^3G_3$
  (middle) partial waves as well as of their coupling amplitude $\epsilon_3$
  (bottom) in the new partial-wave solution. Values are plotted as small solid
  triangles in 10 MeV steps and as small squares in 50 MeV steps together with
  the corresponding total energy $\sqrt s$. The thick solid circle gives
  the energy position of the resonance pole. The dotted curve in the top
  figure is a circle fitted to the loop, its diameter equals to the branching
  ratio $B_{el}$.  
}
\label{fig7}
\end{figure}

A search of the complex energy plane 
revealed a pole in the coupled $^3D_3$-$^3G_3$ wave. Other partial
waves did not change significantly over the energy range spanned by
the new data. 

The fit repeated with different weightings for the new
$A_y$ data resulted in a variation of the pole position and could be
considered a minimal 'error' on its value within the present
fit form. In the weighted fits, a 
pole was located at (2392 - i37) MeV.
The re-fit without weighting produced a pole with 
(2385 - i39) MeV.

\begin{figure}
\centering
\includegraphics[width=0.99\columnwidth]{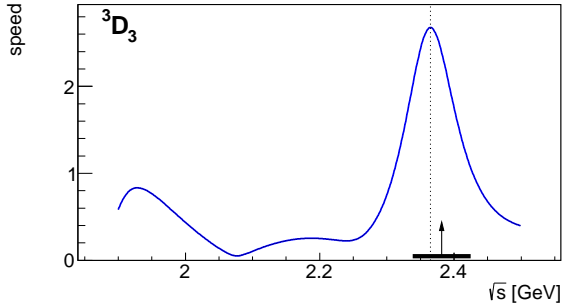}
\caption{\small (Color online) Speed plot of the $^3D_3$ partial wave. The
  dotted vertical line indicates the position of the maximum. Vertical arrow
  and horizontal bar indicate pole and width of the resonance as average over
  the values obtained in contour and speed plots, see text. 
}
\label{fig8}
\end{figure}

For the $^3D_3$ partial wave we display in Fig.~8 a speed plot as defined in
Refs. \cite{Hoehler,hanstein}. It exhibits a
Lorentz-like shape with a maximum at 2.37 GeV and a width of about 80
MeV. Hence
          together with the speed-plot determination we arrive at 
          ($2380\pm10 - i 40\pm5$) MeV as our best estimate for the pole
          position. 

As pointed out by H\"ohler \cite{Hoehler} both Argand diagram and speed plot
allow also the determination of the modulus r of the residue, which corresponds
to just
half of the partial decay width $\Gamma_{el}$ of the resonance into the
elastic channel. Since according to H\"ohler the radius of the resonance
circle in the Argand diagram equals to just half the branching ratio $B_{el} =
\Gamma_{el}/\Gamma$, we  derive from the Argand diagram plotted in Fig.~7, top,
a value of $B_{el}$ = 0.15. The  height $H$ of the Lorentz-like peak above
background in the speed plot is related to the branching
ratio by $H = 2 B_{el} / \Gamma$ , which leads to $B_{el}$ = 0.10. These
numbers depend, of course, somewhat on the assumption of the background. Hence,
as our best estimate for the branching ratio we take the average $B_{el} =
0.12\pm0.03$ (corresponding to r = $5 \pm 1$ MeV). This value agrees
very well to the expectation based on unitarity and the knowledge about the
various two-pion decay channels of the intermediate $\Delta\Delta$ system,
which $d^*$ decays into \cite{PBC}. 

In addition to the modulus r of the residue also its phase $\Phi$ can be
determined - most favorably from the Argand diagram of the derivative of the
partial-wave amplitude \cite{Hoehler, hanstein} - though usually with much
lower accuracy. Here, we obtain values for $\Phi$ between -21 and +9 degrees,
depending on details of the procedure.

\section{Comparison of the New Partial-Wave Solution to Further Observables
  of Neutron-Proton Scattering}

In the following we compare the new SAID partial-wave solution to all data
related to the $np$ scattering issue, which are available for the energy region
of interest. 

\subsection{Total Cross Section}

In Ref. \cite{PBC} the contribution of the $d^*$ resonance to the total $np$
cross section has been estimated to be around 1.5 mb. Though this is small
compared to the total cross section of 38 mb in the resonance region, it is
larger than the uncertainties quoted in the total cross section measurements
of Devlin {\it et al.} \cite{dev}. In fact, the total $np$ cross section data
exhibit a significant rise in this region, whereas the total $pp$ cross section
is flat in the region of interest. 

Fig. 9, top,  shows the total $np$ cross section for $T_n$~=~0.5 - 1.5 GeV. The
data plotted by open squares for $T_n <$ 0.8 GeV are from Lisowski et
al. \cite{lis} taken at LAMPF in a high-resolution dibaryon search. The other
data plotted by open triangles are from Devlin et al. \cite{dev} taken with a
neutron energy resolution of $(4 - 20)\%$ (horizontal bars in Fig.~5). Also data
from Sharov et al. \cite{sha} are shown (open circles), which have larger
uncertainties, but are taken with a much superior neutron energy
resolution of (13 - 15) MeV. The data exhibit a pronounced jump in 
the cross section between $T_n$ = (1.0 - 1.3) GeV. This jump is remarkable,
since 
the $pp$ total cross section is completely flat in this energy region. Hence 
in the isoscalar total nucleon-nucleon cross section $\sigma_{I=0} =
2 \sigma_{pn} - \sigma_{pp}$, where the SAID values are used for
$\sigma_{pp}$, this effect appears still more pronounced (Fig.~9, bottom).   
The current SAID SP07 solution is shown by the solid lines again. Its
description of the data is only fair. In particular the observed 
increase in the total cross section above 1 GeV is only slightly indicated in
the SAID SP07 solution. The dotted lines show the new weighted partial-wave
solution, which includes the resonance pole. They exhibit a clear $s$-shaped
resonance behavior. The dashed lines are averaged over the experimental
resolution of the data of Devlin et al. \cite{dev} and provide a preferable
description of the data. 

\begin{figure} h
\centering
\includegraphics[width=0.99\columnwidth]{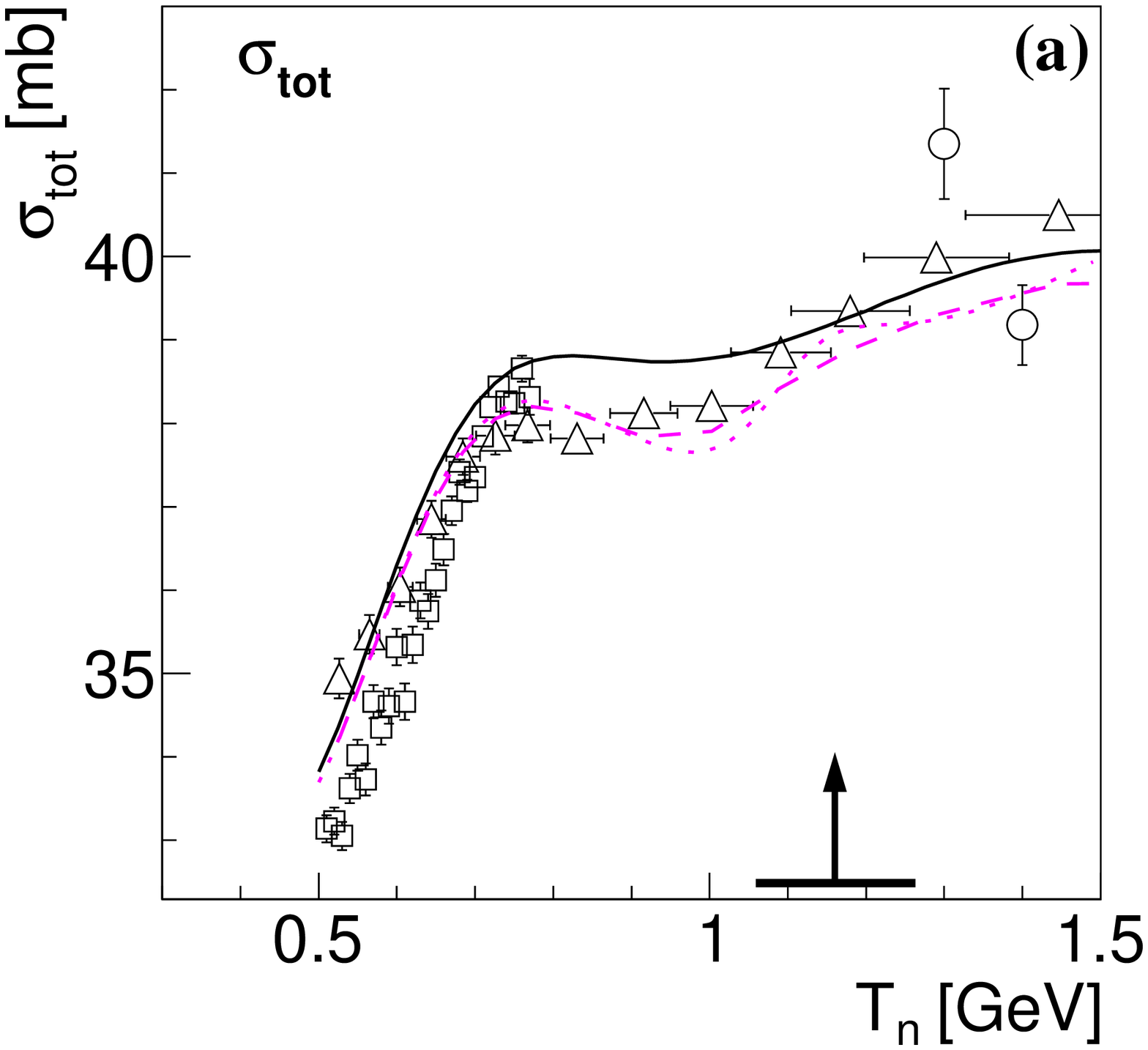}
\includegraphics[width=0.99\columnwidth]{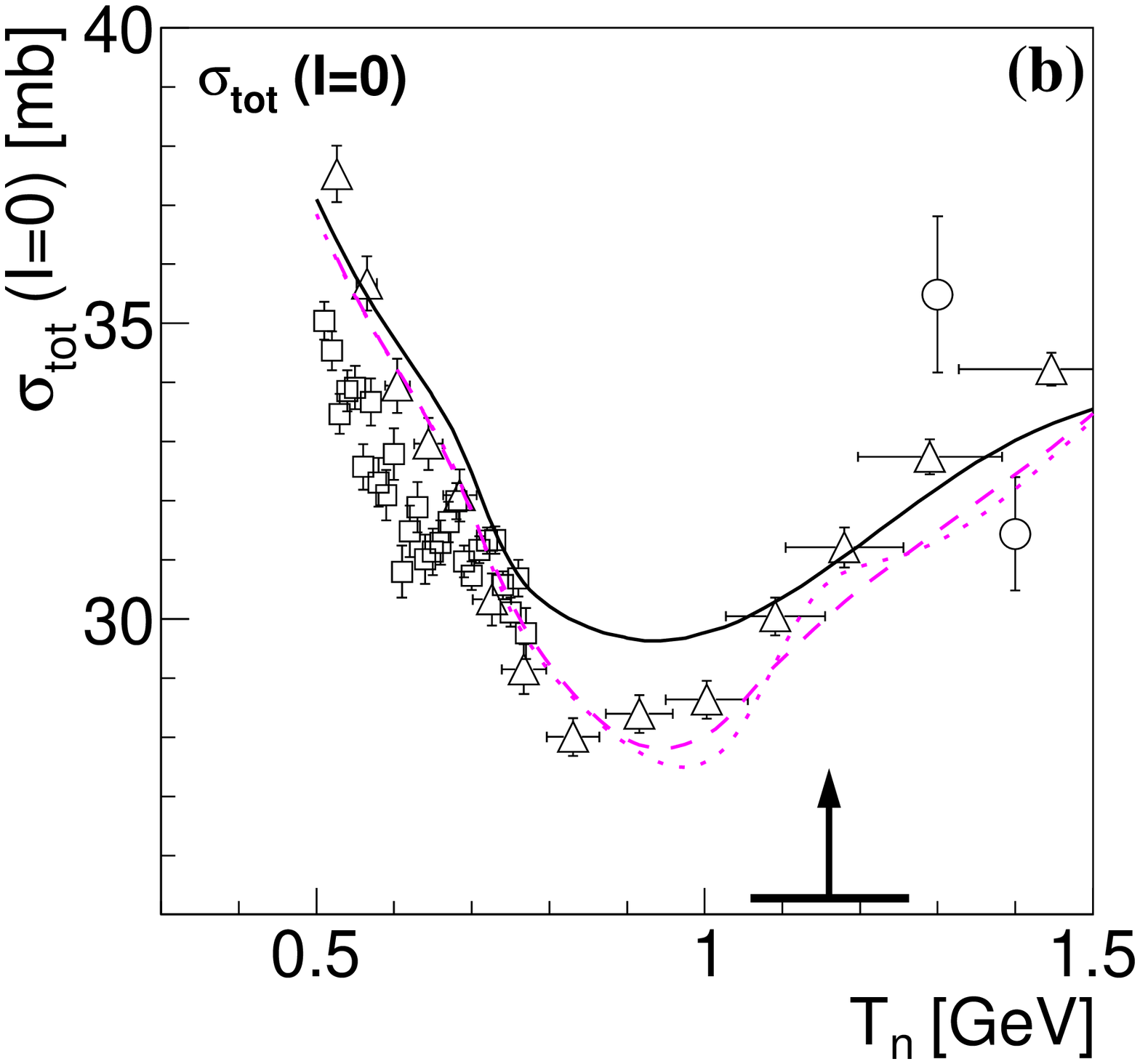}
\caption{\small (Color online) Total pn cross section (top) and total isoscalar
  nucleon-nucleon cross section (bottom) in dependence of the incident neutron
  (nucleon) 
  energy $T_n$. Data are from Lisowski et al. \cite{lis} (open squares),
  Devlin et al. \cite{dev} (open triangles) and Sharov et al. \cite{sha}(open
  circles). The horizontal bars indicate the energy resolution of the incident
  neutrons. Solid and dashed curves representing the SP07 and the new
  solution, respectively, are averaged over the experimental energy
  resolution of Ref. \cite{dev}. The dotted line gives the new solution
  without averaging over 
  the experimental energy resolution. Vertical arrow and horizontal bar
  indicate pole and width of the resonance. 
}
\label{fig9}
\end{figure}

In Fig.~10 we show total $np$ cross sections measured in dependence of the spin
directions of beam and target particles. These spin-dependent total cross
sections are measured with the directions of beam and target polarizations
being either parallel or antiparallel. For the so-called transversal total
cross section difference $\Delta\sigma_T$ the polarization vectors are
transversally oriented with respect to the beam direction. For the so-called
longitudinal cross section difference $\Delta\sigma_L$ they are longitudinally
oriented. 

For $\Delta\sigma_T$ there exist measurements for energies below the resonance
region only, but for $\Delta\sigma_L$  measurements extend to energies well
above this region. Whereas SP07 and new solutions nearly coincide below and in
the resonance region for $\Delta\sigma_T$, they deviate substantially from
each other  
for $\Delta\sigma_L$. On average the new solution gives a superior description
of the $\Delta\sigma_L$ data, in particular in and above the resonance region.

\begin{figure}
\centering
\includegraphics[width=0.99\columnwidth]{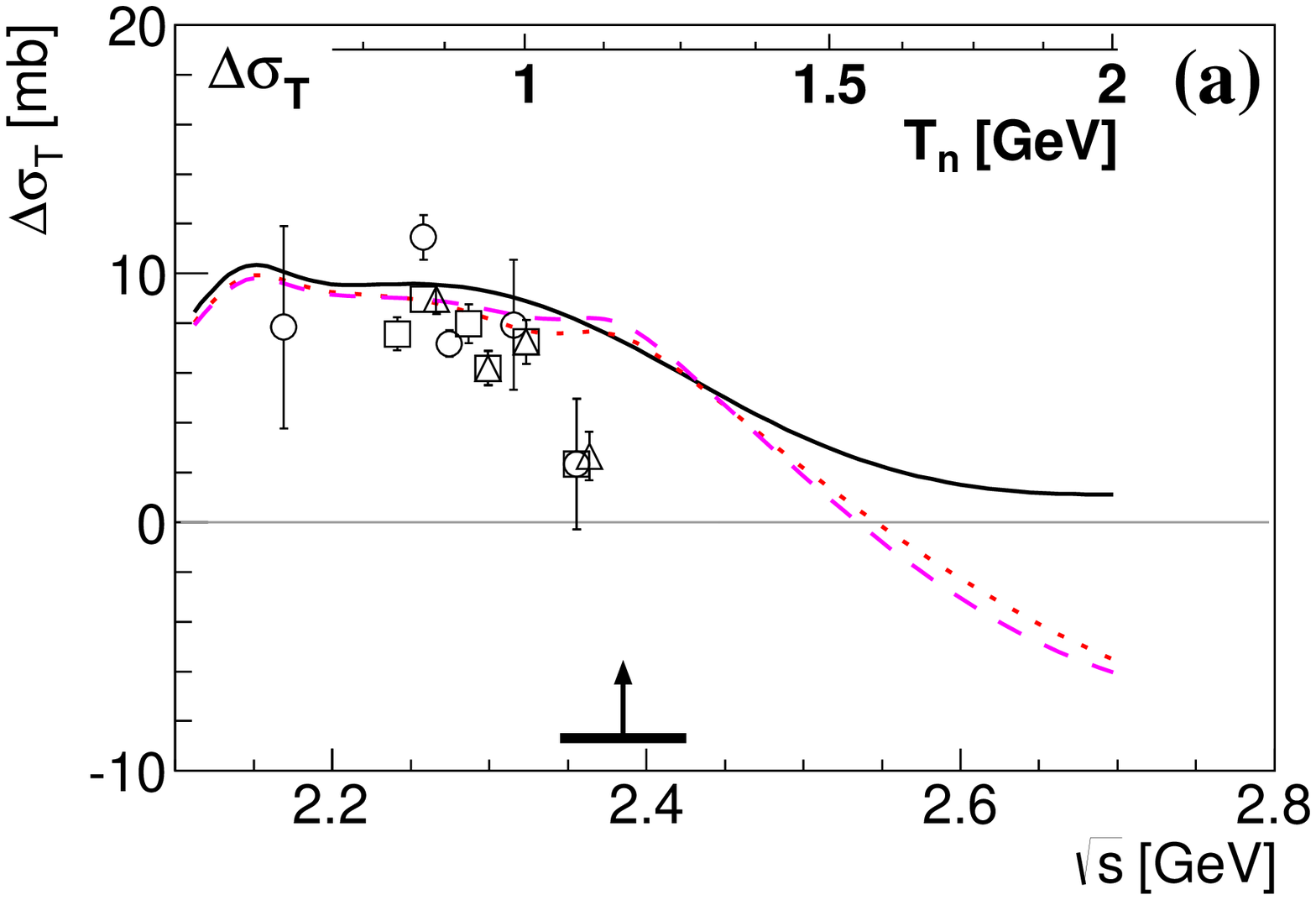}
\includegraphics[width=0.99\columnwidth]{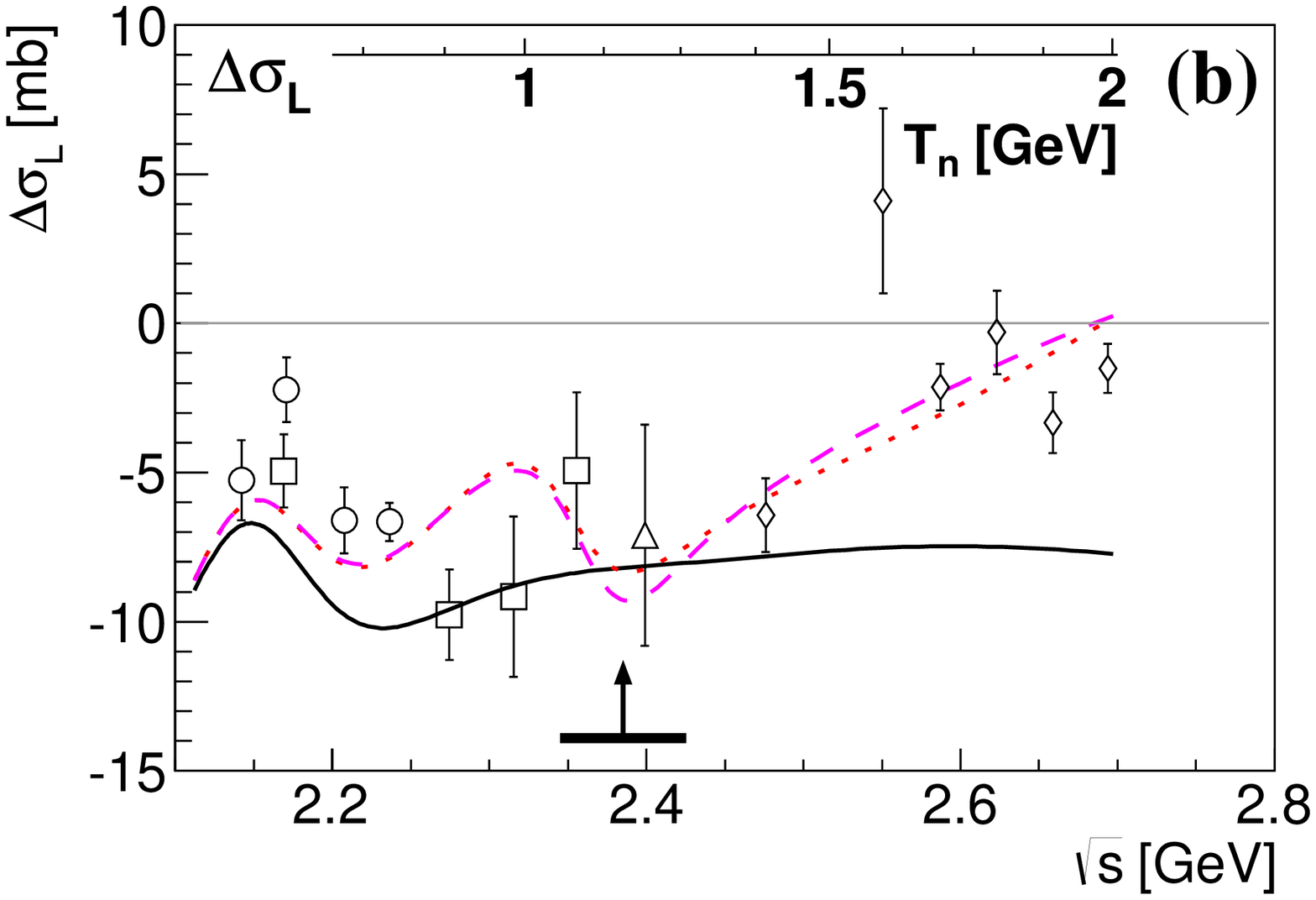}
\caption{\small (Color online) Energy dependence of the transversal (top) and
  longitudinal (bottom) total $np$ cross section differences. The symbols
  denote data from Refs. \cite{les,sha,leh,fon,ball2,bed,adi}. For the
  meaning of the curves see Fig.~4. Vertical arrow and horizontal bar
  indicate pole and width of the resonance.
}
\label{fig10}
\end{figure}

\subsection{Differential Cross Section}

\begin{figure}
\centering
\includegraphics[width=0.99\columnwidth]{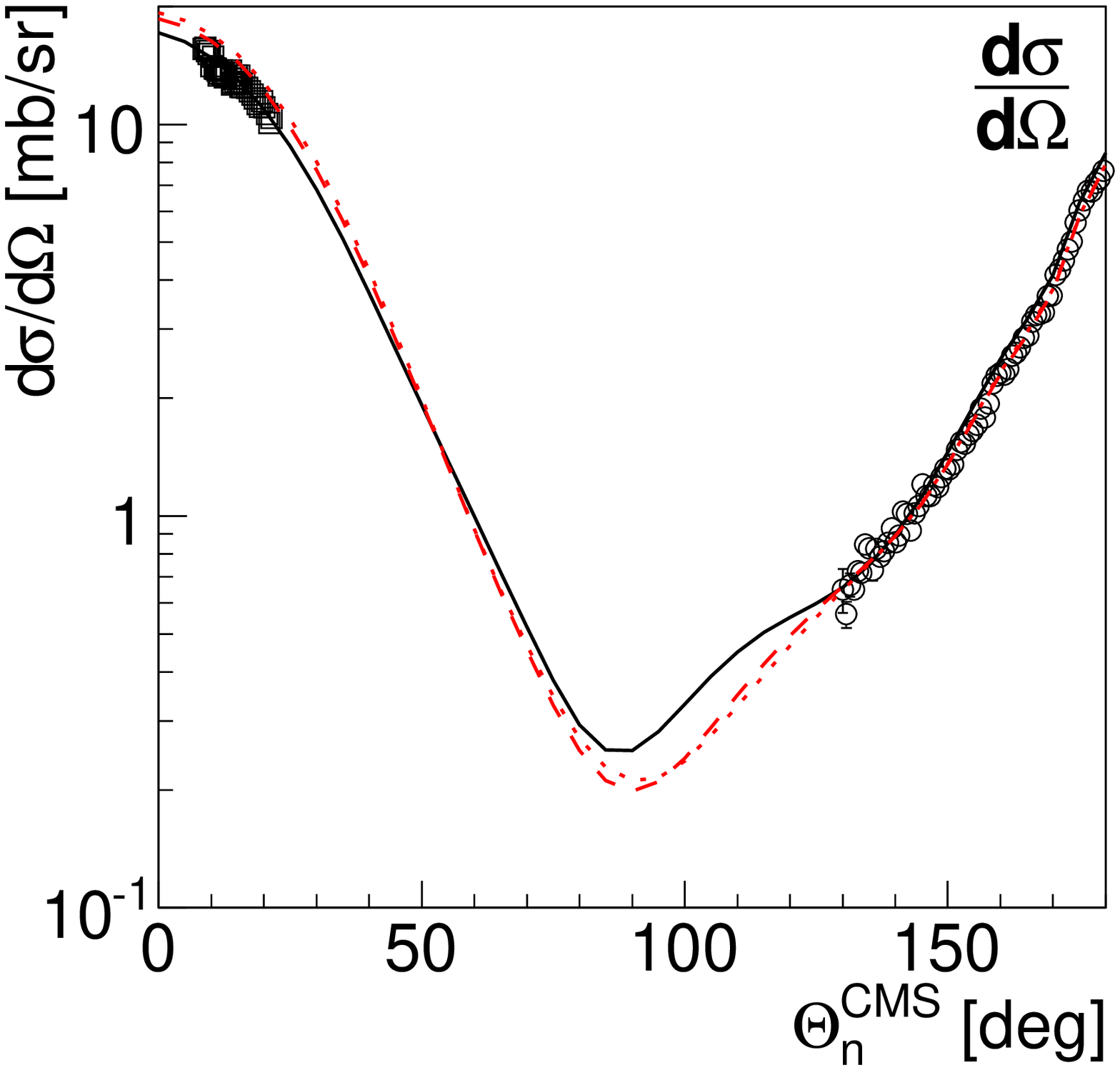}
\caption{\small (Color online) Angular distribution of the differential cross
  section $d\sigma / 
  d\Omega$ at $T_n$ = 1.135 GeV corresponding to the resonance energy $\sqrt
  s$ = 2.38 GeV. For the meaning of the curves see caption of Fig.~4. The
  plotted data are from Ref.  \cite{terrien} ($T_n$ = 1.135 GeV) and Ref.
  \cite{bizard} ($T_n$ = 1.118 GeV).
}
\label{fig11}
\end{figure}

Fig.~11 shows the angular distribution of the differential cross section
$d\sigma / d\Omega$  for elastic $np$ scattering at $T_n$~=~1.135 GeV
corresponding to 
the resonance energy $\sqrt s$~=~2.38 GeV, where we expect the effect of the
resonance on the observables to be largest. At this energy there are only data
for the differential cross section at very forward and backward scattering
angles. The solid line denotes the current SAID SP07 solution, the dashed
(dotted) line gives the result with the new weighted (unweighted) SAID
solution. The resonance effect is small though noticeable in
the differential cross section, since in contrast to the analyzing power the
partial-wave amplitudes enter only quadratically. The resonance effect is
predicted to be most notable at intermediate angles, where the differential
cross section gets smallest and where as of yet no data are available.

\subsection{Spin-Correlation and Spin-Transfer Observables}

As of yet there are no direct measurements of spin-correlation and
spin-transfer measurements in the resonance region. However, 
at COSY-ANKE the reaction $\vec{d}\vec{p} \to n[pp]_s$ has been measured at
various energies \cite{ANKECXX}, where $[pp]_s$ means the proton pair to be in
relative $^1S_0$ state. At ANKE this has been achieved experimentally by
requiring the relative kinetic energy of the two protons to be smaller than 3
MeV.

This reaction is correlated with $np$ scattering in the  
impulse approximation. Whereas the obtained polarization observables agree
very well to calculations based on the SAID SP07 solution for $T_d <$ 2 GeV,
these measurements show large deviations at $T_d$ = 2.27, which corresponds
just to the
resonance energy. A simple relation of the ANKE polarization observables to
specific ones of $np$ scattering is obtained in the $pp$ $^1S_0$ limit at
$\Theta_n^{cm} = 180^\circ$ \cite{LW} for the deuteron-proton tensor analyzing
powers 

\begin{eqnarray}
A_{xx}(q = 0) = &&A_{yy}(q = 0) = \\ \nonumber 
 = &&2~\frac {K_{0ll0}(\pi) - K_{0nn0}(\pi)} {3 - K_{0ll0}(\pi) - 2 K_{0nn0}(\pi)}
\end{eqnarray}

and spin-correlation parameters

\begin{eqnarray}
C_{x,x}(q = 0) = &&C_{y,y}(q = 0) = \\ \nonumber
= &&2~\frac { A_{00nn}(\pi) - D_{n0n0}(\pi)} {3 - K_{0ll0}(\pi) - 2 K_{0nn0}(\pi)},
\end{eqnarray}

where $q$ is the momentum transfer between initial neutron and final proton.
\begin{table} 
\caption{Deuteron-proton tensor analyzing powers and spin-correlation
  parameters at $T_d$ = 2.27 GeV and $q$ = 0 obtained from SAID SP07
  \cite{Arndt07} and new partial-wave solutions 
  by use of eqs. (4) - (5) in comparison with experimental results from
  COSY-ANKE \cite{ANKECXX}.}   
\begin{tabular}{lllll} 
\hline
 & observable&experiment&~~SP07&~new solution\\ 

\hline

& $A_{xx}(0) = A_{yy}(0)$ &~~~-0.38(3) &~~-0.30 &~~~~-0.42 \\

& $C_{x,x}(0) = C_{y,y}(0)$ &~~~-0.39(5) &~~-0.48 &~~~~-0.31 \\


\hline
 \end{tabular}\\
\end{table}
Here $A_{ijkl}$, $D_{ijkl}$ and $K_{ijkl}$ denote neutron-proton spin
correlation and spin-transfer parameters. 

ANKE finds $A_{xx}$ = -0.38(3) and
$C_{x,x}$ = -0.39(5) at $T_d$ = 2.27 GeV (Table 1). Calculating these
observables with the 
SP07 solution results in values -0.30 and -0.48, respectively, which are
significantly different. However, the new SAID solution gives $A_{xx}$ = -0.42
and $C_{x,x}$ = -0.31, which are closer to the ANKE experimental
values.

\section{Summary and Conclusions}

The exclusive and kinematically complete measurements of quasifree polarized
neutron-proton scattering with WASA at COSY have provided detailed
high-statistics data for the analyzing power in the energy range, where a
narrow resonance structure with $I(J^P) = 0(3^+)$, called $d^*$, was observed
previously in the double-pionic fusion to the deuteron. A partial-wave
analysis including these new data exhibits a resonance pole at ($2380\pm10 - i
40\pm5$) MeV in the $^3D_3$-$^3G_3$ coupled partial waves -- establishing thus
the $d^*(2380)$ resonance structure to be a genuine $s$-channel
resonance. This constitutes the first clear-cut experimental finding of a true
dibaryon resonance. 

This resonance has been first observed in the reaction $pn \to d\pi^0\pi^0$
reaction. The Dalitz plot there \cite{MB} shows that the resonance
predominantly decays via an intermediate $\Delta\Delta$ configuration. Exactly
such a state with identical quantum numbers has been predicted first by Dyson
and Xuang \cite{dyson} based on SU(6) symmetry breaking already in 1964 -- just
shortly after Gell-Mann's publication of the quark model
\cite{gellmann}. Whereas this dibaryon state was denoted by $D_{03}$ in
Ref. \cite{dyson}, it was named $d^*$ later-on by Goldman {\it et.al.}
\cite{goldman}, who 
pointed out the unique symmetry properties of such a state with these quantum
numbers calling it the "inevitable" dibaryon.  

So $d^*(2380)$ may be associated with a bound $\Delta\Delta$ resonance, which
could contain a mixture of asymptotic $\Delta\Delta$ and six-quark, hidden
color configurations \cite{BBC}. Recent quark-model calculations
\cite{ping,zhang,Huang} 
find this state at a mass close to the experimental one. Whereas the
width calculated in Ref.~\cite{ping} is still substantially too large, the one
obtained in Ref.~\cite{Huang} is already in good agreement with the
experimental finding, as soon as coupling to hidden color configurations is
accounted for. New three-body calculations
\cite{GG1,GG2} of Faddeev type with relativistic kinematics and hadron dynamics
find $d^*(2380)$ at the right mass and only slightly larger width \footnote{In
  Refs. \cite{GG1,GG2} a width suppression factor has been introduced, which is
  at variance with our observations for the $d^* \to pp\pi^0\pi^-$ decay
  \cite{TS}. Hence only the values for the width without this factor should be
  compared to the experimental one.}.

In addition to $np$ scattering evidence for $d^*$ has been found so far in the
two-pion production reactions $pn \to d\pi^0\pi^0$, $pn \to d\pi^+\pi^-$ and
$pn \to pp\pi^0\pi^-$. So the only remaining hadronic channels, where $d^*$
should contribute, are $np\pi^0\pi^0$ and $np\pi^+\pi^-$. The first one has
been studied at WASA and is on the way of being 
published. The latter channel has been measured at HADES and preliminary
results have been reported already at conferences. This means that all major
channels of the $d^*$ decay will have been investigated in near future. All
decay branches into the two-pion decay channels appear to be well accounted
for by isospin relations \cite{PBC}. 

It is amazing that the simple but basic estimate of Dyson and Xuong
\cite{dyson} is already very close to the now observed mass of $d^*$ alias
$D_{03}$. Together with the results of the new calculations based on quark
\cite{ping,zhang,Huang} and/or hadron dynamics \cite{GG1,GG2} this might give
some confidence in the predictive power of these theoretical considerations on
further dibaryon states. 

We finally note that very recently an alternative explanation of the WASA data
has been suggested by D. V. Bugg \cite{Bugg}. Among the issuses raised in this
paper we just mention here one. The specific intermediate $NN^*(1440)$ system
proposed for the production of the narrow resonance-like structure
is not restricted to the isoscalar channel and hence should show up both in
$np$ and $pp$ initiated two-pion production. However, as we have shown, this
narrow resonance structure is missing in all $pp$-initiated two-pion
production channels including double-pionic fusion
\cite{isofus,FK,deldel,nnpipi,iso}.  

\section{Acknowledgments}

We acknowledge valuable discussions with J. Haidenbauer, Ch. Hanhart,
A. Kacharava and C. Wil\-kin on this issue. 
This work has been supported by BMBF, Forschungszentrum J\"ulich
(COSY-FFE), the U.S. Department of Energy Grant DE-FG02-99ER41110, 
the Polish National Science Centre (grant No. 2011/03/B/ST2/01847) and
the Foundation for Polish Science (MPD).

\end{document}